\documentclass[]{aastex631}

\usepackage{amsmath}
\begin{document}

\title{Entropy-mode imprints in the solar corona: non-exponential damping and phase shifts of compressive oscillations} 

\correspondingauthor{Dmitrii Kolotkov}
\email{D.Kolotkov.1@warwick.ac.uk}

\author[0000-0002-0687-6172]{Dmitrii Kolotkov}
\affiliation{Centre for Fusion, Space and Astrophysics, Department of Physics, University of Warwick,\\ Coventry, CV4 7AL, UK}
\affiliation{Engineering Research Institute \lq\lq Ventspils International Radio Astronomy Centre (VIRAC)\rq\rq\ of Ventspils University of Applied Sciences,\\ Inzenieru iela 101, Ventspils, LV-3601, Latvia}

\author[0000-0002-3505-9542]{Sergey Belov}
\affiliation{Centre for Fusion, Space and Astrophysics, Department of Physics, University of Warwick,\\ Coventry, CV4 7AL, UK}

\author[0009-0001-9903-5179]{Mohamed Sherif}
\affiliation{Mullard Space Science Laboratory, University College London,\\ Holmbury St. Mary, Dorking, Surrey, RH5 6NT, UK}

\begin{abstract}
Magnetohydrodynamic (MHD) waves in coronal loops provide key seismological diagnostics through their characteristic time signatures. While fast and slow magnetoacoustic modes are routinely exploited, the entropy mode, despite being another eigenmode of the system, remains largely inaccessible due to its non-propagating and non-oscillatory nature.
We identify {possible} observable time-domain signatures of the entropy mode and its indirect effects. Our approach exploits the intrinsically non-adiabatic conditions of the solar corona, under which the entropy mode is closely linked to the compressive slow mode. We consider a one-dimensional coronal loop model with field-aligned thermal conduction, where standing slow and entropy modes are simultaneously excited.
We show that the entropy mode leaves distinct imprints on the total loop temperature {{and density}} perturbations. Specifically, its rapid decay relative to the slow mode produces a non-exponential damping profile during the initial oscillation cycles and introduces a pronounced asymmetry between the upper and lower temperature {{and density}} envelopes. These effects arise naturally from the superposition of two exponentially decaying components with different damping timescales. Furthermore, deviations from the canonical quarter-period phase shift between temperature{{/density}} and velocity perturbations in the standing slow mode are explained by the entropy-mode effect.
We conclude that the entropy mode {may be} detected through its impact on compressive oscillations. Revealing its role in non-exponential damping, envelope asymmetry, and phase shifts of compressive oscillations {makes the entropy mode potentially accessible to observations} and lays the foundation for solar and stellar seismological applications.
\end{abstract}


\section{Introduction} \label{sec:intro}

Magnetohydrodynamic (MHD) waves in the corona of the Sun are open to direct multi-wavelength studies with decent spatial, temporal and spectral resolution \citep{2020ARA&A..58..441N}. They offer a reliable tool for remote diagnostics of otherwise inaccessible local physical conditions \citep{2024RvMPP...8...19N}, crucial for constraining and improving space weather models \citep{2015AdSpR..55.2745S}, and are considered to play an important role in the transport and conversion of magnetic energy into heat to maintain multi-million K temperature of coronal plasmas \citep{2020SSRv..216..140V}. On other stars, MHD waves and oscillations are observed indirectly, as quasi-periodic pulsations in flare lightcurves \citep{2021SSRv..217...66Z}, which provide an interesting opportunity for studying solar-stellar analogies and probing conditions in the atmospheres of distant stars and their planetary systems.

In the time domain, the application of MHD coronal seismology is based on identifying the characteristic time signatures of the wave processes in theory and observations, and linking them to the underlying physical mechanisms. For example, Gaussian damping profile of transverse kink-mode oscillations of coronal loops \citep[e.g.][]{2012A&A...539A..37P, 2013A&A...551A..39H, 2016A&A...589A.136P} and an apparent linear scaling between their oscillation period and damping time \citep[e.g.][]{2002SoPh..206...99A, 2002ApJ...577..475R, 2002A&A...394L..39G, 2006RSPTA.364..433G, 2013A&A...552A.138V} are used to probe the process of resonant absorption via mode coupling to Alfv\'en continuum and fine transverse structuring of coronal loops. The existence of two distinct decaying and decayless regimes of kink oscillations and the transition between them over time indicate their role in sustaining the coronal energy balance \citep[e.g.][]{2013A&A...552A..57N, 2023MNRAS.525.5033Z, 2024MNRAS.531.4611N}. The scaling of kink oscillation damping time with amplitude is used as a signature of nonlinear effects \citep[e.g.][]{2016A&A...590L...5G, 2016A&A...595A..81M, 2021ApJ...910...58V}. Deviations of the kink overtone period ratio $P_1/nP_n$ from unity are connected to the effects of coronal density stratification, axial and/or perpendicular density structuring, and provide information about the corresponding spatial scales \citep[e.g.][]{2009SSRv..149....3A, 2009A&A...493..259I, 2013ApJ...777...17S, 2017ApJ...842...99L, 2018ApJ...854L...5D, 2019A&A...632A..64D}. The characteristic tadpole- and boomerang-shaped structures in wavelet spectra of dispersively evolving fast magnetoacoustic wave trains enable their detection in high-cadence observations with limited spatial resolution \citep[e.g.][]{2004MNRAS.349..705N, 2011SoPh..273..393M, 2020SSRv..216..136L, 2021MNRAS.505.3505K}.

Time-domain MHD seismology with slow waves similarly relies on their characteristic temporal signatures, which encode information on the magnetic field, plasma density, and thermodynamic processes of coronal plasmas. Thus, the phase shifts between plasma temperature, density, and velocity perturbations in a slow wave allow one to distinguish between standing \citep[also known as SUMER oscillations, with quarter-period phase shift between Doppler shift and intensity perturbations, see e.g.][]{2002ApJ...580L..85O, 2003A&A...406.1105W, 2003A&A...402L..17W, 2006ApJ...639..484M, 2015ApJ...807...98Y} and propagating \citep[with in-phase dynamics, e.g.][]{2002SoPh..209..265S} slow waves. Deviations from those expected values indicate effects of non-ideal processes, such as thermal conduction, viscosity, heating/cooling misbalance \citep[e.g.][]{2009A&A...494..339O, 2011ApJ...727L..32V, 2015ApJ...811L..13W, 2019MNRAS.483.5499K, 2021SoPh..296..105P, 2022SoPh..297....5P, 2023FrASS..1067781Z}. Although phase shifts in slow waves are commonly attributed to non-ideal effects, their precise physical origin remains an open question. Similarly to kink modes, slow-mode waves exhibit frequency-dependent damping, with scaling between damping time (length) and period being distinctly different in active region loops \citep{2008ApJ...685.1286V, 2011SSRv..158..397W, 2016ApJ...820...13M, 2019ApJ...874L...1N, 2022MNRAS.514L..51K, 2023A&A...677A..23A} and in polar plume/interplume regions \citep[e.g.][]{2014ApJ...789..118K, 2018ApJ...853..134M}, which remains unexplained. \citet{2016ApJ...830..110C} detected an analogous scaling between the oscillation period and damping time of SUMER-type quasi-periodic pulsations in solar and stellar X-ray flares, which indicated the similarity of the flare mechanisms on the Sun and other stars. Time delays between quasi-periodic pulsations in extreme ultraviolet and soft X-ray flare emissions, potentially caused by standing fast sausage mode and standing slow mode, indicate effects of magnetic trapping of accelerated electrons and/or cooling of the thermal plasma \citep{2012ApJ...749L..16D}. As in kink oscillations, the nonlinear damping of slow waves is manifested through scaling of the damping time with oscillation amplitude \citep[e.g.][]{2008ApJ...685.1286V, 2013A&A...553A..23R, 2019ApJ...874L...1N}.
There is also evidence of slow-wave phase speed scaling with period \citep{2025MNRAS.538..797Z}, attributed to the slow-wave dispersion effect \citep[waveguide or non-adiabatic,][]{2021SoPh..296..122B, Zavershinskii2026}.

Apparent damping profiles of {propagating} slow waves were theoretically found not to obey the exponential law due to the non-uniform multi-thermal transverse structure of coronal loops and line-of-sight integration effects \citep{2024A&A...683A.109V, 2024ApJ...970...58K, 2024SoPh..299....2F}. Even in the absence of transverse non-uniformities, preliminary signatures of {standing} slow waves exhibiting non-exponential damping during the first few oscillation cycles were found in \citet{2022FrASS...973664K}, a behaviour that also appears (but is not discussed) in observations \citep[see e.g. Fig.~2 in][]{2011SSRv..158..397W}. In \citet{2022FrASS...973664K}, this has been broadly associated with the rapid decay of another MHD eigenmode of the coronal loop, the entropy mode, although no specific implications of this association were identified.

In ideal MHD, the entropy mode is a non-oscillatory and non-propagating solution (with the real and imaginary parts of the oscillation frequency $\omega_\mathrm{R}=0$ and $\omega_\mathrm{I}\ne 0$, respectively), which remains decoupled from other modes and is often neglected in the analysis \citep[see e.g. the non-propagating initial perturbation marked by $I$ in][]{2014ApJ...788...44M}. \citet{2011A&A...533A..18M} suggested that cool, dense blobs caused by the entropy mode in the vicinity of magnetic null points can be used for indirect observations of nanoflares in the solar corona. Episodes of catastrophic cooling, formation of prominences and coronal rain are also implicitly associated with the entropy mode instability \citep[see e.g.][and references therein]{2022FrASS...920116A}. In intrinsically non-adiabatic conditions of the solar corona, entropy and slow magnetoacoustic modes of a coronal loop were shown to be strongly related and may possess mixed properties \citep{2021SoPh..296...96Z, 2023FrASS..1067781Z, 2023Physi...5..193K}. In other words, in the presence of non-adiabatic processes in the plasma, such as thermal conduction, optically thin radiation, free magnetic energy release and conversion into heat (all are intrinsic for solar and stellar coronae), external plasma perturbations excite both a slow-mode wave and an entropy mode, the dynamics and time evolution of which are closely connected. The characteristic time signatures of this connection and their potential seismological implications remain largely unexploited.

In this Letter, we present characteristic time signatures of the entropy mode {that may aid its} identification in observations through its link to the slow-mode wave.
We demonstrate that the simultaneous development of a slow mode and an entropy mode in the coronal plasma with field-aligned thermal conduction leads to a non-exponential damping of the total signal. We also explain the mechanism for the commonly observed phase shift between plasma temperature{{/density}} and velocity perturbations in a standing slow wave to deviate from the expected $\pi/2$ (quarter-period) value by the effect of the entropy mode development. The paper is structured as follows. In Section~\ref{sec:dispersion}, we describe the governing equations and derive and compare the damping time of the slow and entropy modes through the dispersion relation analysis. In Section~\ref{sec:numeric}, we present how the entropy-mode effect appears in the total standing perturbation of the plasma temperature {{and density}} obtained numerically; suggest practical steps for estimating entropy mode damping time and amplitude from it (Section~\ref{sec:damping_time}); derive a link between the entropy-mode amplitude and the slow-mode phase shifts between plasma temperature{{, density}} and velocity perturbations (Section~\ref{sec:shifts}); demonstrate the spatial structure of the entropy mode along the loop and investigate its excitation effectivness in a broad range of equilibrium plasma temperatures and densities (Section~\ref{sec:spatial}). Our final thoughts, perspectives, and conclusions are summarised in Section~\ref{sec:concl}.

\section{Dispersion relation analysis} \label{sec:dispersion}

In strongly magnetised coronal plasma conditions, the evolution of predominantly longitudinal standing SUMER-type oscillations and field-aligned propagating EUV disturbances can be effectively modelled within the 1D (infinite magnetic field) approximation \citep[see e.g.][and references therein]{2009SSRv..149...65D, 2011SSRv..158..267B, 2021SSRv..217...76B, 2016GMS...216..419B, 2011SSRv..158..397W, 2021SSRv..217...34W}, using governing equations (linearised),
\begin{align}\label{eq:motion}
&\rho_0 \frac{\partial u}{\partial t} = -\frac{\partial p}{\partial z},\\
&\frac{\partial \rho}{\partial t} +  \rho_0\frac{\partial u}{\partial z} = 0,\\
&p=\frac{k_\mathrm{B}}{m}\left( \rho_0 T + T_0 \rho\right),\\
&\frac{\partial {T}}{\partial t} - (\gamma-1)\frac{T_0}{\rho_0}\frac{\partial {\rho}}{\partial t}=\frac{\kappa_\parallel}{\rho_0 C_\mathrm{V}}\frac{\partial^2{T}}{\partial z^2}.\label{eq:energy}
\end{align}
Similar 1D models are used to describe plasma dynamics along coronal loop strands \citep[e.g.][]{2014LRSP...11....4R}. 
In this framework, the wave propagation direction is prescribed by the direction of the local magnetic field ($z$-axis), and the effects of wave-induced perturbations of the infinitely stiff magnetic field are neglected. The variables $u$, $T$, $\rho$, and $p$ in Eqs.~(\ref{eq:motion})--(\ref{eq:energy}) stand for the perturbations of the field-aligned plasma velocity, temperature, mass density, and thermal pressure, respectively. The equilibrium state is described by $T_0$, $\rho_0$, and $p_0$, with $u_0=0$ (no flows). The right-hand side of the energy equation (\ref{eq:energy}) accounts for the effect of field-aligned heat conduction, as the dominant damping mechanism of plasma perturbations along the field \citep[see e.g.][]{2003A&A...408..755D, 2011ApJ...727L..32V, 2015ApJ...811L..13W, 2025A&A...693A.186B}.
{More specifically, \citet{2021SSRv..217...34W} showed that the thermal conduction is dominant over other mechanisms in the damping of compressive oscillations for the typical hot coronal loops with $\rho_0\approx10^{-12}$--$10^{-11}$\,kg\,m$^{-3}$, $T_0\approx5$--10\,MK, and $L\gtrsim 100$\,Mm.}
In this work, we use the standard local Spitzer form of the heat conduction coefficient $\kappa_\parallel = 10^{-11}T_0^{5/2}\,\mathrm{W\,m}^{-1}\,\mathrm{K}^{-1}$ \citep[see][for comparison of local and non-local heat conduction models]{2023FrASS..1055124A}. The constant parameters $m$, $k_\mathrm{B}$, $\gamma$, and $C_\mathrm{V} = (\gamma-1)^{-1}k_\mathrm{B}/m$ represent the mean coronal particle mass (0.6 of proton mass), the Boltzmann constant, the adiabatic index (\(5/3\)), and the standard specific heat capacity at constant volume, respectively. {The use of a linearised form of Eqs.~(\ref{eq:motion})--(\ref{eq:energy}) is justified by the numerical work of \citet{2018ApJ...860..107W} who showed that the development of nonlinearity in standing slow-magnetoacoustic waves is effectively suppressed by the effect of viscosity \citep[see also the discussion of this point in][Sec.~7]{2021SSRv..217...34W}. In addition, according to numerical simulations performed in, e.g., \citet{2004ApJ...605..493M, 2007SoPh..246..187S, 2008ApJ...685.1286V, 2015ApJ...813...33F}, and analytical modelling in \citet{2013A&A...553A..23R, 2025MNRAS.542.1076R}, the shape of the oscillation signal gets highly anharmonic in the nonlinear regime, which is inconsistent with the AIA and SUMER observations of compressive oscillations.}

Seeking solution to Eqs.~(\ref{eq:motion})--(\ref{eq:energy}) in a harmonic form $\exp{i(k z - \omega t)}$ yields the dispersion relation linking the perturbation's angular frequency $\omega$ and the wavenumber $k$ \citep[see e.g.][]{2003A&A...408..755D, 2014ApJ...789..118K, 2016ApJ...820...13M},
\begin{equation}\label{eq:disp}
    \omega^3 + A(k)\omega^2 + B(k)\omega + C(k) = 0,
\end{equation}
where
\begin{equation}
    A(k) = \frac{i}{\tau_\mathrm{cond}},
    B(k) = -c_\mathrm{s}^2k^2,
    C(k) = -\frac{ic_\mathrm{s}^2k^2}{\gamma \tau_\mathrm{cond}},
    \nonumber
\end{equation}
with the characteristic thermal conduction time $\tau_\mathrm{cond}= \rho_0 C_\mathrm{V} k^{-2}/\kappa_\parallel$ and standard sound speed $c_\mathrm{s} = \sqrt{\gamma k_\mathrm{B}  T_0 / m}$.

Without thermal conduction ($\tau_\mathrm{cond}\to \infty$), Eq.~(\ref{eq:disp}) is second-order in $\omega$ and describes two slow magnetoacoustic waves propagating in opposite directions along the $z$-axis at a constant speed $c_\mathrm{s}$. Including the finite effect of thermal conduction, Eq.~(\ref{eq:disp}) becomes third-order in \(\omega\), yielding two propagating slow-mode waves coupled to a non-propagating entropy (thermal) mode. Focusing on a standing wave form with real wavenumber $k=\pi/L$ (prescribed by the coronal loop length $L$) and complex frequency $\omega=\omega_\mathrm{R}+i\omega_\mathrm{I}$, Eq.~(\ref{eq:disp}) can be solved for the slow mode approximately (under the assumptions $\omega\tau_\mathrm{cond}\gg 1$ and $\omega_\mathrm{I}\ll\omega_\mathrm{R}$):
\begin{align}
    &\omega_\mathrm{R} = c_\mathrm{s}k,\\
    &\omega_\mathrm{I} = \frac{1}{2}\frac{\gamma-1}{\gamma}\frac{1}{\tau_\mathrm{cond}},
\end{align}
and for the entropy mode:
\begin{align}
    &\omega_\mathrm{R} = 0,\\
    &\omega_\mathrm{I} = \frac{1}{\gamma \tau_\mathrm{cond}}.
\end{align}
The obtained solutions for $\omega_\mathrm{I}$ of the slow and entropy modes provide corresponding damping times,
\begin{align}\label{eq:dampt-slow}
    &\tau_\mathrm{slow} = \frac{2\gamma}{\gamma-1}\tau_\mathrm{cond},\\
    &\tau_\mathrm{entropy} = \gamma \tau_\mathrm{cond},\label{eq:dampt-ent}
\end{align}
and their ratio $\tau_\mathrm{slow}/\tau_\mathrm{entropy}=3$ for $\gamma=5/3$. In other words, as long as the effect of thermal conduction remains weak ($\omega\tau_\mathrm{cond}\gg 1$), the entropy mode in a coronal loop decays three times faster than the slow mode.

\begin{figure}
	\centering
        \includegraphics[width=0.49\linewidth]{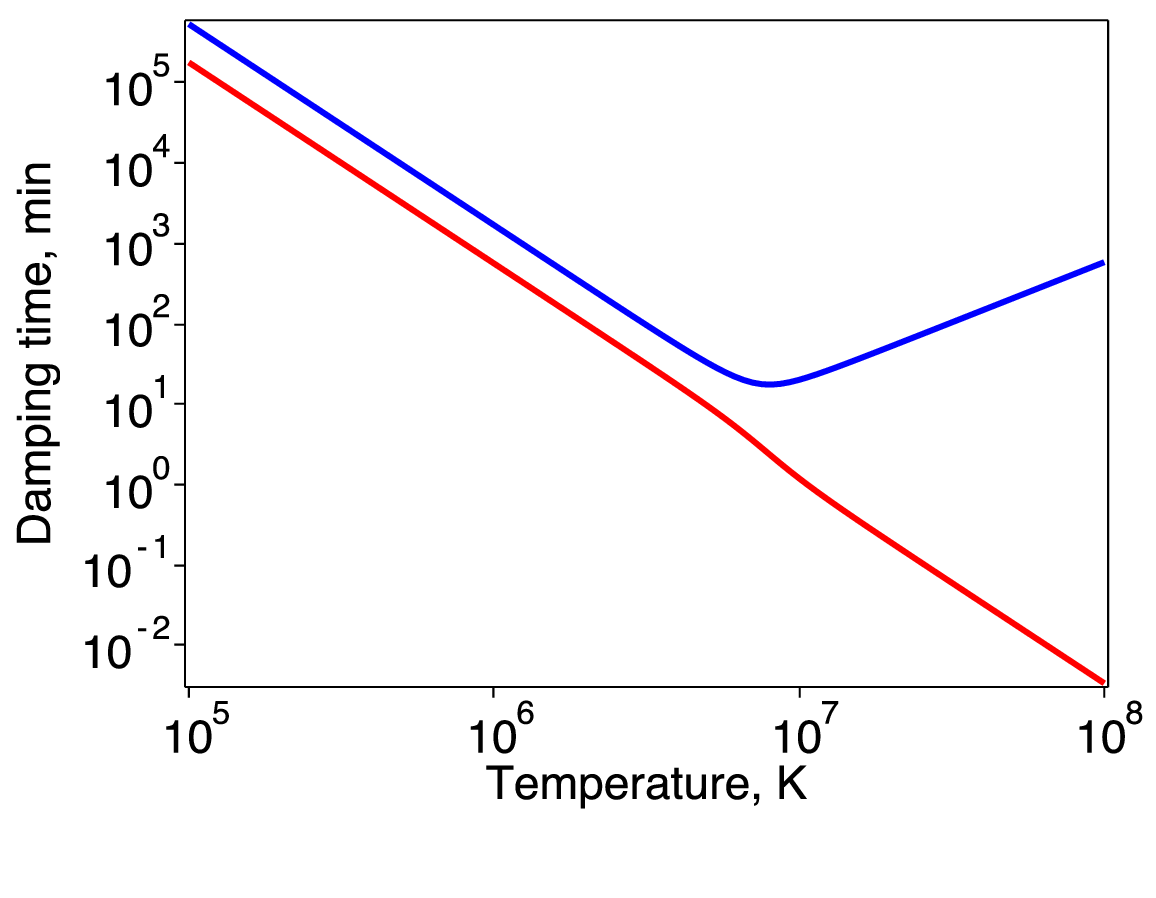}
        \includegraphics[width=0.49\linewidth]{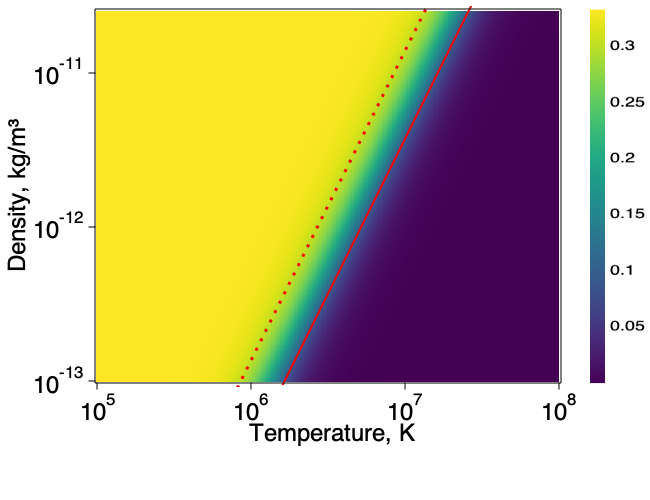}
	\caption{Left: the damping time of the standing slow, $\tau_\mathrm{slow}$ (blue) and entropy, $\tau_\mathrm{entropy}$ (red) modes, obtained from a numerical solution of Eq.~(\ref{eq:disp}) as a function of temperature for a loop of length $L=180$\,Mm and plasma density $\rho_0 = 3\times10^{-12}$\,kg\,m$^{-3}$.
    Right: ratio $\tau_\mathrm{entropy}/\tau_\mathrm{slow}$ from Eq.~(\ref{eq:disp}) as a function of loop temperature and density (for the same loop length). Contours of $\tau_\mathrm{entropy}/\tau_\mathrm{slow} = 0.3$ and $0.1$ are shown as red dotted and solid lines, respectively.
	}
	\label{fig:damping_slow_entropy}
\end{figure}

To obtain slow- and entropy-mode damping times with arbitrary thermal conduction, we solve the dispersion relation given by Eq.~(\ref{eq:disp}) numerically for the loop length $L=180$\,Mm and a broad range of plasma temperature $T_0$ (0.1--100\,MK) and density $\rho_0$ ($10^{-13}$--$10^{-10}$\,kg\,m$^{-3}$). As shown in Fig.~\ref{fig:damping_slow_entropy}, the ratio of damping times $\tau_\mathrm{slow}/\tau_\mathrm{entropy}$ remains around 3 until thermal conduction becomes over-effective and changes the plasma evolution to an isothermal regime ($\tau_\mathrm{slow}\to\infty$). For example, for $\rho_0=3\times10^{-12}$\,kg\,m$^{-3}$, this happens at around $T_0=10$\,MK. Thus, we conclude that for all plausible combinations of coronal plasma density and temperature and a typical loop length, the entropy mode decays at least three times faster than the slow standing mode.

\section{Numerical analysis} \label{sec:numeric}

\subsection{Damping times and amplitudes of slow and entropy modes} \label{sec:damping_time}

Based on the fact that the entropy mode decays substantially faster than the slow mode for all considered combinations of coronal plasma temperature and density (Sec.~\ref {sec:dispersion}), in this Section, we propose steps for deducing the entropy-mode damping time and amplitude in practice, from the time-series analysis of oscillatory plasma perturbations. For this, we solve Eqs.~(\ref{eq:motion})--(\ref{eq:energy}) numerically in a resonator centered at $z=0$, with the \emph{pdsolve} function in the mathematical environment Maple 2024.2. The \emph{pdsolve/numeric} routine uses a second-order (in space and time) centred, implicit finite difference scheme, the applicability of which to the type of Eqs.~(\ref{eq:motion})--(\ref{eq:energy}) was discussed in detail, including the numerical accuracy test, in \citet{2022FrASS...973664K}.
The plasma perturbations $T$, $\rho$, and $u$ are normalised to $T_0=6.3$\,MK (typical for SUMER observations), $\rho_0=5\times10^{-12}$\,kg\,m$^{-3}$, and $c_\mathrm{s} \mathrm{[km/s]} \approx 152 \sqrt{T_0\mathrm{[MK]}}=381$\,km\,s$^{-1}$, respectively.
We use the wavelength $\lambda_0=360$\,Mm of the fundamental standing harmonic (prescribed by the loop length $\lambda_0=2L$) and the adiabatic acoustic wave oscillation period $P_0 = \lambda_0/c_\mathrm{s} = 15.7$\,min to normalise spatial and time variables in Eqs.~(\ref{eq:motion})--(\ref{eq:energy}). As coronal slow waves are well known observationally and theoretically to reflect efficiently from the lower atmosphere \citep[due to the strong acoustic impedance mismatch, see e.g.][]{2013ApJ...779L...7K, 2015ApJ...804....4K, 2015ApJ...813...33F, 2016ApJ...826L..20R}, we implement fully reflective rigid-wall boundary conditions $u(\pm\lambda_0/4,t)=0$, $\partial_z\rho(\pm\lambda_0/4,t)=0$. For the initial condition, we adopt $T(z,0)=0$, $\rho(z,0)=0$, and $u(z,0)=0.1\cos(2\pi z/\lambda_0)$ \citep[i.e. the wave is excited by the injected flow, cf.][]{2012ApJ...754..111O}, which corresponds to the fundamental harmonic excitation to exclude distortions to the initial oscillation cycles by transition processes \citep[cf. Figs.~1 and 4 in][respectively]{2005A&A...436..701S, 2019A&A...628A.133K}.

\begin{figure}
	\centering
        \includegraphics[width=0.5\linewidth]{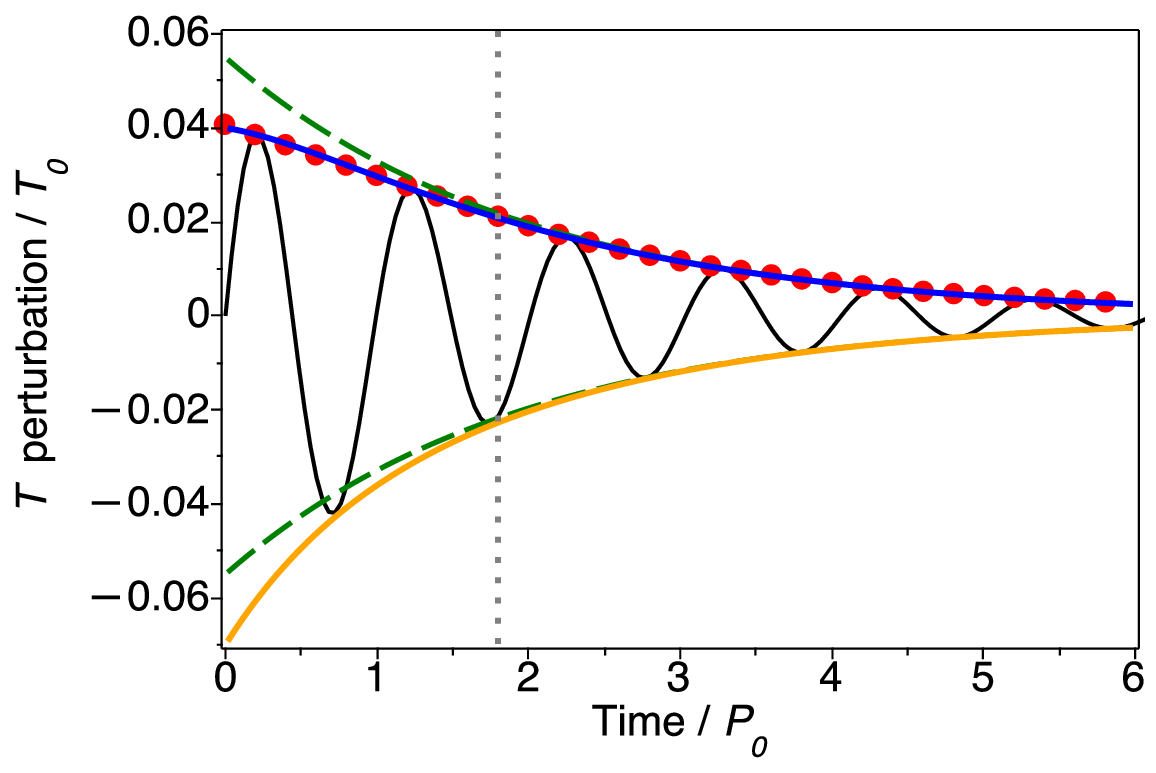}
        \includegraphics[width=0.42\linewidth]{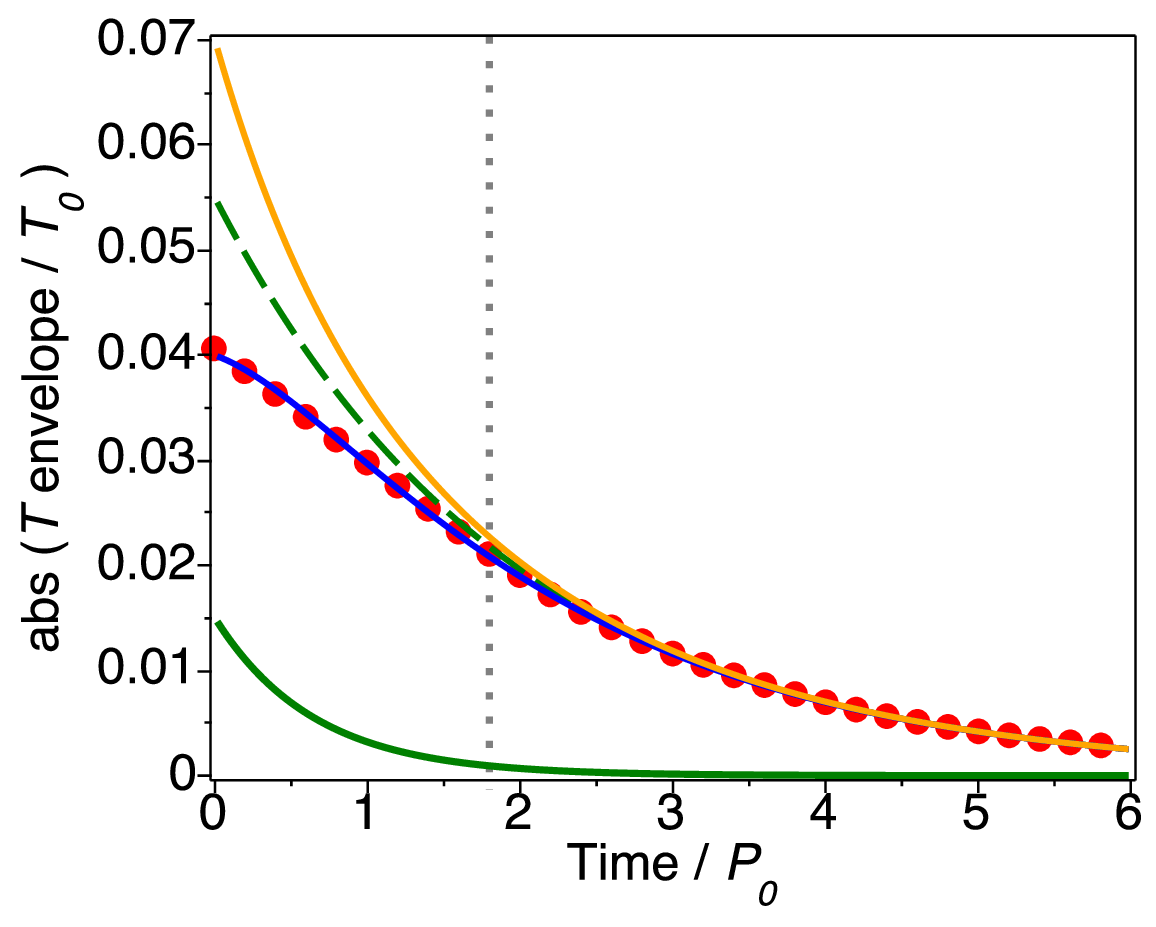}
	\caption{Left: relative loop temperature perturbation (black), obtained from the numerical solution of Eqs.~(\ref{eq:motion})--(\ref{eq:energy}) at $z_0=0.15$ (normalised to the wavelength, $\lambda_0=2L$ for $L=180$\,Mm). The signal's upper and lower envelopes are shown with the blue line (and red circles) and the orange line, respectively. The green dashed lines show exponential best-fits of the signal's upper and lower envelopes at $t>3\tau_\mathrm{entropy}$.
    Right: temperature perturbation envelopes (blue line with red circles and orange line, both shown as absolute values), decomposed into two exponentially decaying components: the slow mode (green dashed) and the entropy mode (green solid).
    Accordingly, the sum and difference of the green dashed and green solid lines reproduce the orange (lower envelope) and blue (upper envelope) lines, respectively.
    The vertical dotted line in both panels indicates $3\tau_\mathrm{entropy}$ given by Eq.~(\ref{eq:dampt-ent}).
	}
	\label{fig:envelopes}
\end{figure}

The left-hand panel of Fig.~\ref{fig:envelopes} shows the oscillatory solution for the plasma temperature perturbation $T(t)$ that we obtain from the numerical setup described above, at $z_0=0.15\lambda_0$. Using cubic spline interpolation of the signal's local extrema, we construct its upper and lower envelopes (see the blue and orange lines in Fig.~\ref{fig:envelopes}). Since our simulation is fully linear (Eqs.(\ref{eq:motion})--(\ref{eq:energy}) are linearised) and 1D, we do not expect any modifications to the signal's damping law caused by nonlinearity \citep[][]{2008ApJ...685.1286V, 2013A&A...553A..23R} or loop's transverse multi-thermality \citep{2024A&A...683A.109V, 2024SoPh..299....2F}. Despite this, in Fig.~\ref{fig:envelopes}, we observe a clear departure of the $T$ perturbation envelopes from the exponential shape during the first few oscillation cycles, which we attribute to the effect of simultaneous development of the entropy and slow magnetoacoustic modes in the resonator. Both modes decay exponentially in full agreement with the linear analysis, but their superposition results in a distinctly non-exponential decay of the total signal. As predicted by the dispersion analysis (Sec.~\ref {sec:dispersion}), the entropy mode decays faster than the slow mode, hence the total $T$ perturbation signal in Fig.~\ref{fig:envelopes} returns to the usual exponential damping pattern for $t \gtrsim 3\tau_\mathrm{entropy}$ given by Eq.~(\ref{eq:dampt-ent}) which serves as an effective damping regimes switch time \citep[cf.][for kink oscillations]{2017A&A...600A..78P}, when the entropy mode disappears.

We can extract the slow-mode damping time and initial amplitude from the total $T$ perturbation signal by best-fitting either of its envelopes with a single decaying exponential function $A_{T,\mathrm{slow}}^\mathrm{fit}\exp(-t/\tau^\mathrm{fit}_\mathrm{slow})$ at $t > 3\tau_\mathrm{entropy}$, i.e. skipping the initial oscillation cycles where the envelopes are distorted from exponential by the entropy mode. We perform this best-fitting in Maple, using the \emph{NonlinearFit} function which allows for a local numeric solution to both linear and nonlinear least-squares problems. It provides $A_{T,\mathrm{slow}}^\mathrm{fit} = 0.054\pm 0.0003$ and $\tau^\mathrm{fit}_\mathrm{slow}=1.95\pm 0.002$ (both are in normalised units) with 3-$\sigma$ uncertainties, and the resulting slow-mode exponential envelope is shown in Fig.~\ref{fig:envelopes} (green, dashed). We can account for the effect of the entropy mode and estimate its damping time and initial amplitude by refining the best-fitting model of the total $T$ perturbation envelope as
\begin{equation}\label{eq:slow_ent_model}
    T_\mathrm{env}(t)= A_{T,\mathrm{entropy}}^\mathrm{fit}\exp\Biggl(-\frac{t}{\tau^\mathrm{fit}_\mathrm{entropy}}\Biggr) + A_{T,\mathrm{slow}}^\mathrm{fit}\exp\Biggl(-\frac{t}{\tau^\mathrm{fit}_\mathrm{slow}}\Biggr),
\end{equation}
and taking the values of the slow-mode damping time $\tau^\mathrm{fit}_\mathrm{slow}$ and initial amplitude $A_{T,\mathrm{slow}}^\mathrm{fit}$ obtained at the previous step. Best-fitting the total $T$ signal's upper envelope by model~(\ref{eq:slow_ent_model}) with the \emph{NonlinearFit} function in Maple gives $A_{T,\mathrm{entropy}}^\mathrm{fit} = -0.014\pm 0.0003$ and $\tau^\mathrm{fit}_\mathrm{entropy}=0.6\pm 0.024$ (normalised), indicating that the entropy mode acts to suppress positive and enhance negative $T$ perturbations in the resonator, resulting into asymmetric upper and lower envelopes of the total $T$ perturbation signals (see the right-hand panel in Fig.~\ref{fig:envelopes}). The derived values of $\tau^\mathrm{fit}_\mathrm{slow}\approx 1.95$ and $\tau^\mathrm{fit}_\mathrm{entropy}\approx 0.6$ give ratio $\tau^\mathrm{fit}_\mathrm{slow}/\tau^\mathrm{fit}_\mathrm{entropy} = 3.25$ which is approximately consistent with the prediction of the dispersion analysis (Sec.~\ref{sec:dispersion}), while the mismatch with the expected value of 3 can be attributed to best-fitting uncertainty and deviation of the slow-mode evolution from the weakly non-adiabatic regime for the considered equilibrium plasma parameters $T_0=6.3$\,MK and $\rho_0=5\times10^{-12}$\,kg\,m$^{-3}$.

\subsection{Link to slow-mode phase shifts} \label{sec:shifts}

\begin{figure}
    \includegraphics[width=0.49\linewidth]{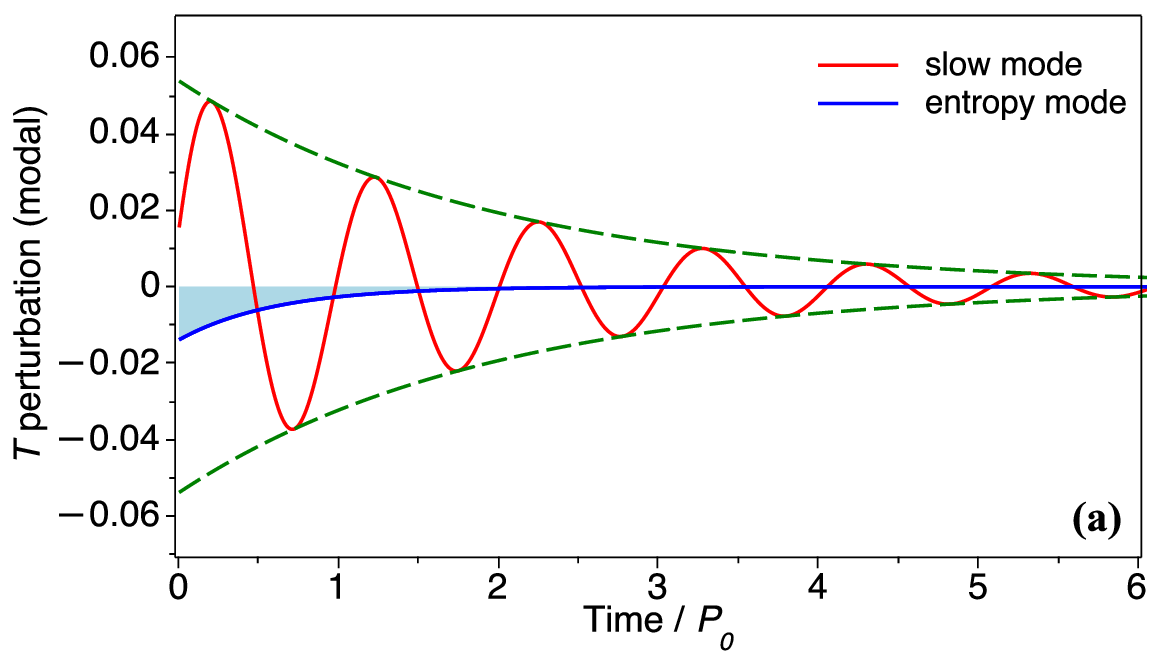}
     \includegraphics[width=0.5\linewidth]{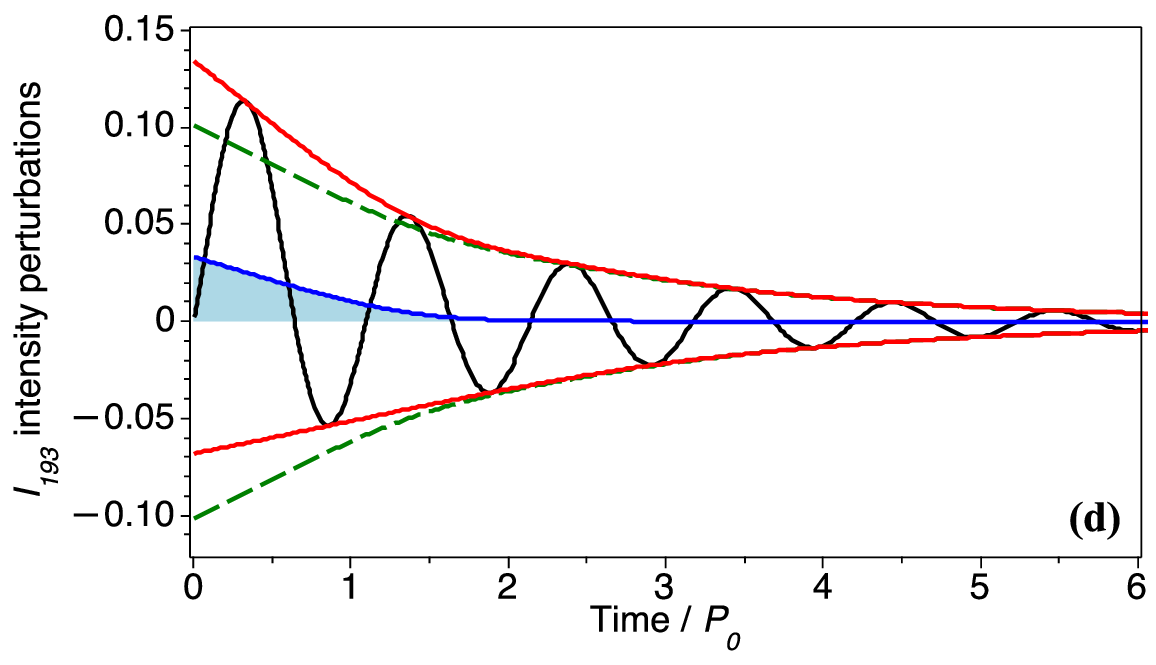}
        \includegraphics[width=0.49\linewidth]{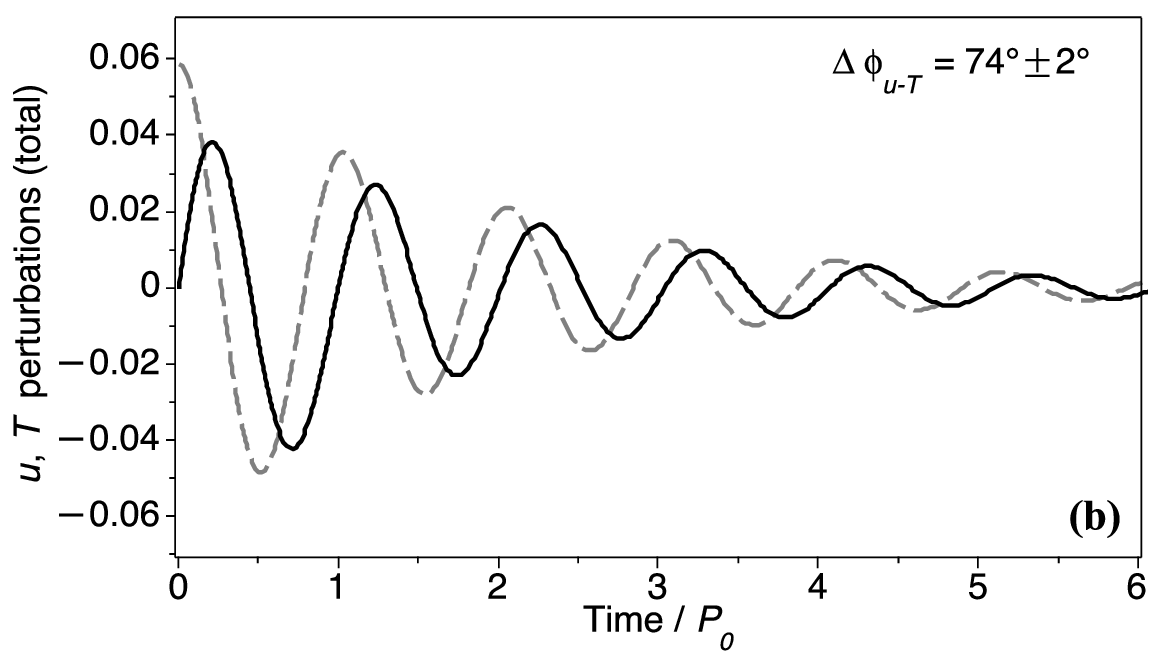}
        \includegraphics[width=0.5\linewidth]{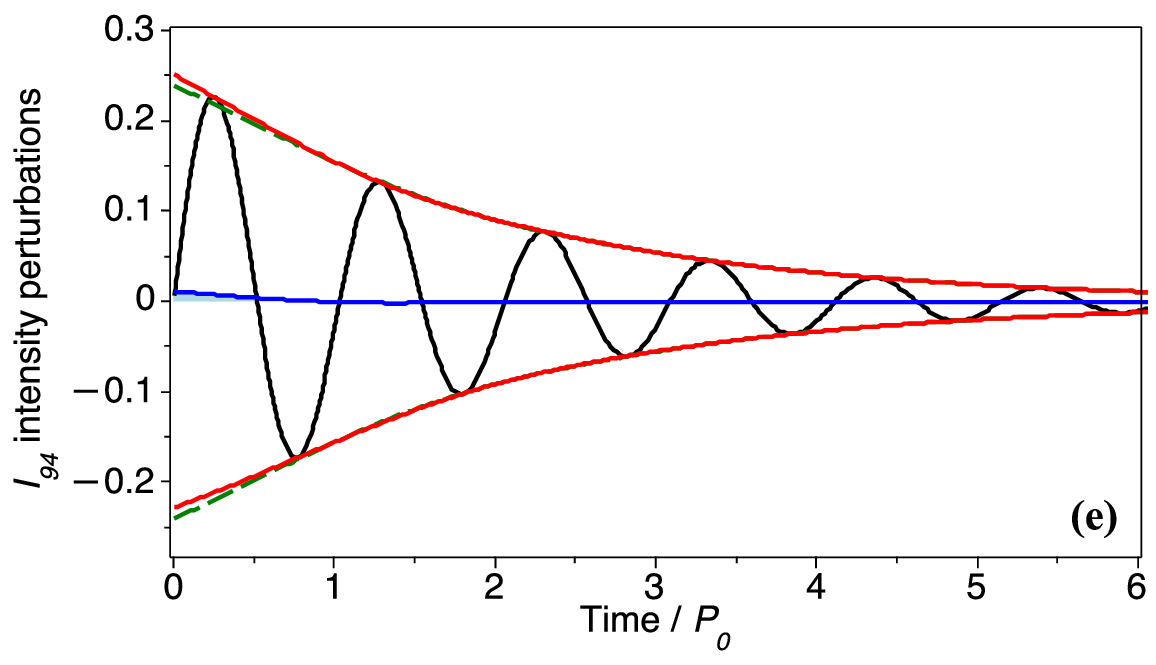}
        \includegraphics[width=0.49\linewidth]{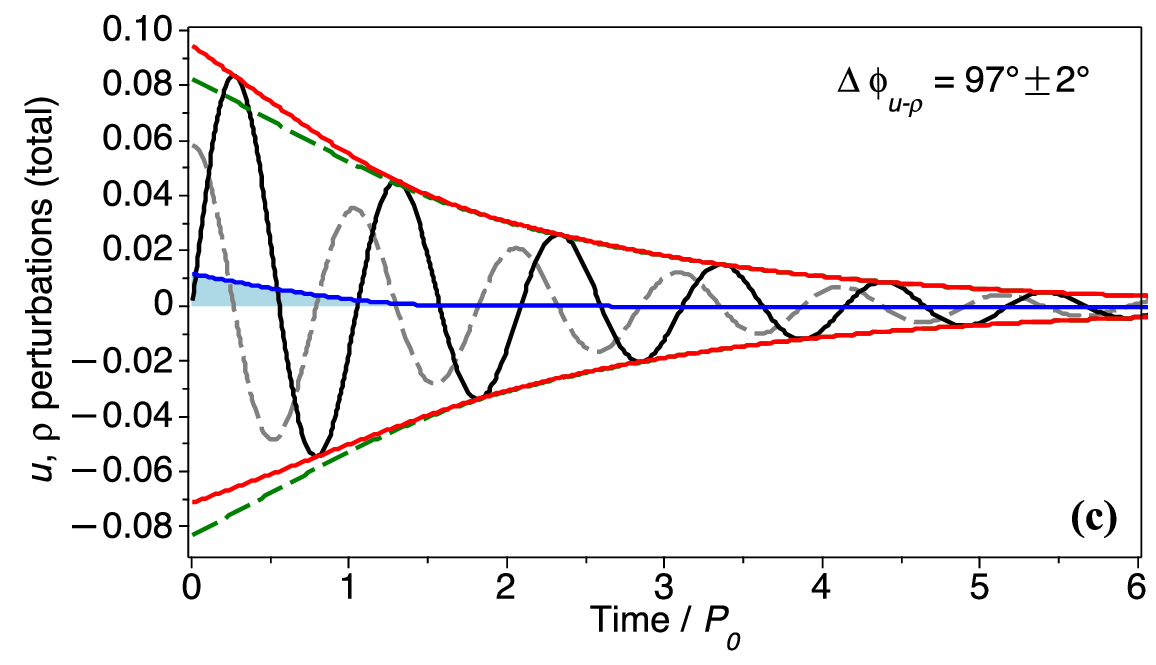}
	\caption{Panel (a): time evolution of the slow-mode component (red) and the entropy-mode component (blue) given by Eqs.~(\ref{eq:entropy_sol})--(\ref{eq:slow_sol}) in the total loop temperature perturbation shown in Fig.~\ref{fig:envelopes} (and in panel (b)). Green dashed lines show exponentially decaying slow-mode envelopes.
    Panel (b): total loop temperature perturbation (black solid, same as in Fig.~\ref{fig:envelopes}) and velocity perturbation (grey dashed), obtained from numerical solution of Eqs.~(\ref{eq:motion})--(\ref{eq:energy}) at $z=0.15\lambda_0$.
    The blue-shaded region in panel (a) highlights the amplitude mismatch between the oscillatory signals shown in panel (a) in red and panel (b) in black solid.
    {{Panel (c): total density perturbation (solid black) obtained numerically and its upper and lower envelopes (solid red) decomposed into the slow-mode component (green dashed) and the entropy-mode component (blue solid with shading). As in panel (b), the grey dashed line shows the velocity perturbation.}}
    {Panel (d): intensity perturbations in the 193\,\AA\ channel of SDO/AIA, synthesised as described in Section~\ref{sec:shifts}. The total envelopes, their slow-mode and entropy-mode components are shown in red, dash green, and blue shading, respectively.}
    {Panel (e): same as panel (d), but for the 94\,\AA\ channel.}
	}
	\label{fig:shifts}
\end{figure}

For model (\ref{eq:slow_ent_model}) to satisfy initial conditions $T(z,0)=0$ used to obtain the temperature perturbation shown in Fig.~\ref{fig:envelopes}, one needs to account for the phase behaviour of the slow mode (the standing entropy mode has zero phase). Thus, we can decompose the total $T$ perturbation shown in Fig.~\ref{fig:envelopes} (at fixed $z$) into a non-oscillatory entropy-mode component and an oscillatory slow-mode component \citep{2023FrASS..1067781Z}, both exponentially decaying, as
\begin{align}\label{eq:entropy_sol}
    &T_\mathrm{entropy}(t) = A_{T,\mathrm{entropy}}\exp\Biggl(-\frac{t}{\tau_\mathrm{entropy}}\Biggr),\\
    &T_\mathrm{slow}(t) = A_{T,\mathrm{slow}}\exp\Biggl(-\frac{t}{\tau_\mathrm{slow}}\Biggr)\cos\left(\frac{2\pi}{P}t-\Delta\phi_\mathrm{ideal}+\Delta\phi_\mathrm{diss}\right).\label{eq:slow_sol}
\end{align}
The slow-mode oscillation period $P$ in Eq.~(\ref{eq:slow_sol}) can be estimated from the Fourier analysis of the total temperature perturbation signal shown in Fig.~\ref{fig:envelopes} as $P\approx1.025$ (relative to the adiabatic acoustic value $P_0=15.7$\,min) which is consistent with the conductive slow-wave dispersion effect \citep[see e.g. Fig.~2 in][]{2022FrASS...973664K}.
The parameters $\Delta\phi_\mathrm{ideal}$ and $\Delta\phi_\mathrm{diss}$ describe phase shifts between plasma temperature and velocity perturbations in the considered wave. Thus, $\Delta\phi_\mathrm{ideal}$ is well known to be caused by the wave's standing nature and is equal to $\pi/2$ \citep[i.e. quarter-period, see e.g.][]{2002ApJ...580L..85O, 2003A&A...406.1105W, 2003A&A...402L..17W, 2006ApJ...639..484M}. In turn, $\Delta\phi_\mathrm{diss}$ appears due to non-ideal, dissipative effects, such as thermal conduction \citep[e.g.][]{2009A&A...494..339O, 2011ApJ...727L..32V, 2015ApJ...811L..13W, 2022FrASS...973664K, 2023FrASS..1067781Z}. Whilst observations of $\Delta\phi_\mathrm{diss}$ in standing or propagating compressive waves are extensively used to probe thermodynamic properties of the coronal plasma, the exact physical mechanism of its generation, i.e. what causes temperature perturbations to de-tune from the velocity and/or density in the non-ideal coronal plasma, remains uncertain. Our analysis allows for linking the appearance of $\Delta\phi_\mathrm{diss}$ in the slow mode with the development of the entropy mode. Indeed, demanding total temperature perturbation $T_\mathrm{entropy} + T_\mathrm{slow}$ given by Eqs.~(\ref{eq:entropy_sol})--(\ref{eq:slow_sol}) to be zero at $t=0$ to satisfy the initial condition $T(z,0)=0$, one arrives to the following relationship,
\begin{equation}
    \Delta\phi_\mathrm{diss}(u,T) =\arcsin\left(-\frac{A_{T,\mathrm{entropy}}}{A_{T,\mathrm{slow}}}\right).\label{eq:shift_diss}
\end{equation}
In other words, the decrease of temperature in the resonator caused by the entropy mode speeds up the development of the temperature perturbation in the slow mode, thus reducing its phase delay with respect to the velocity perturbation from the ideal $\pi/2$ (quarter-period) value. In our case, using the values of $A_{T,\mathrm{entropy}}=-0.014$ and $A_{T,\mathrm{slow}}=0.054$ obtained from the best-fitting in Sec.~(\ref{sec:damping_time}), we obtain $\Delta\phi_\mathrm{diss}(u,T)\approx0.26$\,rad. Together with the ideal geometrically-induced shift of $-\pi/2$, it results in a total phase shift of the slow-mode temperature perturbation vs. velocity perturbation of $(-\pi/2 + 0.26)\times180^{\circ}/\pi\approx-75^{\circ}$. The entropy-mode and slow-mode temperature perturbations determined by Eqs.~(\ref{eq:entropy_sol})--(\ref{eq:slow_sol}) with $\Delta\phi_\mathrm{diss}$ given by Eq.~(\ref{eq:shift_diss}) are shown in the top panel of Fig.~\ref{fig:shifts}, the superposition of which reproduces the total temperature perturbation shown in Fig.~\ref{fig:envelopes}.

From our numerical solution of Eqs.~(\ref{eq:motion})--(\ref{eq:energy}), we can explicitly obtain the time evolution of the plasma velocity perturbation $u(t)$ at a given coordinate $z_0=0.15\lambda_0$ (see panel (b) in Fig.~\ref{fig:shifts}). Applying cross-correlation analysis between $u(t)$ and $T(t)$ signals from the numerical solution yields a time lag $\Delta t(u,T) = 0.21\pm0.005$ (with uncertainty taken at the numerical time step as a conservative estimate), providing $[\Delta t(u,T) / P ]\times 360^{\circ}\approx 74^{\circ} \pm 2^{\circ}$ which agrees well with our phase shift estimation via Eq.~(\ref{eq:shift_diss}) and the entropy-mode amplitude.

{{We can also apply our analysis to the numerical solution for the density perturbation (panel (c) in Fig.~\ref{fig:shifts}). Namely, we obtain the upper and lower density envelopes (see the red lines), the asymmetry between which indicates the presence of the entropy-mode effect. As the entropy mode is found to modify the signal's upper and lower envelopes by the same amount (see Section~\ref{sec:damping_time}), the mean of the envelopes gives us the time evolution of the entropy mode (see the blue line in panel (c) of Fig.~\ref{fig:shifts}), from which we obtain the entropy-mode amplitude in the density perturbation, $A_{\rho,\mathrm{entropy}}\approx 0.01$ (relative to $\rho_0$). Subtracting the entropy signal (blue line) from the total density perturbation's upper envelope (red line), we obtain the time evolution of the slow-mode envelope (green dashed), providing the slow-mode amplitude in the density perturbation, $A_{\rho,\mathrm{slow}}\approx 0.08$ (also relative to $\rho_0$). Adopting Eq.~(\ref{eq:shift_diss}) for $A_{\rho,\mathrm{entropy}}$ and $A_{\rho,\mathrm{slow}}$ as $\Delta\phi_\mathrm{diss} (u,\rho) =\arcsin(-{A_{\rho,\mathrm{entropy}}}/{A_{\rho,\mathrm{slow}}})$, we obtain the dissipative component of the phase shift between density and velocity perturbations $\Delta\phi_\mathrm{diss} (u,\rho)\approx -0.13$\,rad, resulting in the total shift of $(-\pi/2 - 0.13)\times180^{\circ}/\pi\approx-97^{\circ}$. This matches well the value of the time lag between density and velocity perturbations $\Delta t(u,\rho) = 0.275\pm0.005$ and the corresponding phase shift $[\Delta t(u,\rho) / P ]\times 360^{\circ}\approx 97^{\circ} \pm 2^{\circ}$, estimated from the cross-correlation analysis.
}}

{To illustrate the combined entropy-mode effect of plasma density and temperature perturbations on the observed intensity $I(t)$, we consider an optically thin regime in which $I\propto \rho^2 G(T)$, with $G(T)$ being the instrument response function. In the linear regime, this reduces to $\delta I/I_0 = 2\delta\rho/\rho_0 + G_\lambda \delta T/T_0$, where $G_\lambda = T_0/G_0 \times dG_0/dT_0$ is the observational wavelength dependent coefficient. For example, using the 193\,\AA\ and 94\,\AA\ channels of SDO/AIA \citep[e.g.][]{2017ApJ...847L...5P, 2021ApJ...914...81K}, we obtain $G_{193} \approx -1.62$ and $G_{94} \approx 1.63$ for the considered $T_0 = 6.3$\,MK \citep[e.g.][]{2012SoPh..275...41B, 2014SoPh..289.2377B}. Thus, as the entropy mode is found to act in anti-phase on the temperature and density perturbations (see panels (a) and (c) in Fig.~\ref{fig:shifts}), the envelope asymmetry enhances in the 193\,\AA\ intensity $I_{193}$ and suppresses almost to zero in the 94\,\AA\ intensity $I_{94}$, as illustrated in panels (d) and (e), respectively.} 

\subsection{Spatial structure and excitation effectiveness of the standing entropy mode} \label{sec:spatial}

\begin{figure}
	\centering
        \includegraphics[width=0.4\linewidth]{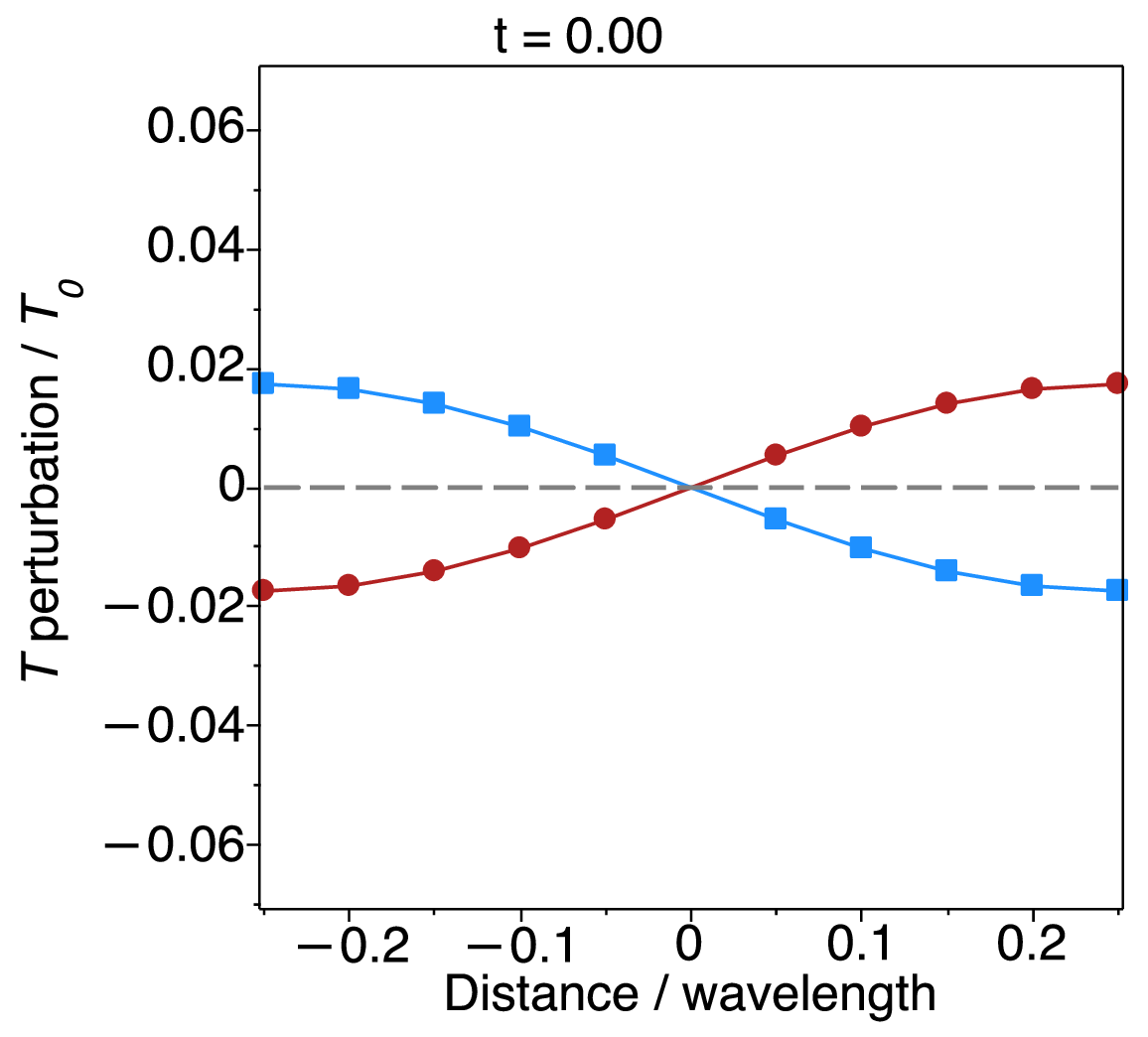}
        \includegraphics[width=0.4\linewidth]{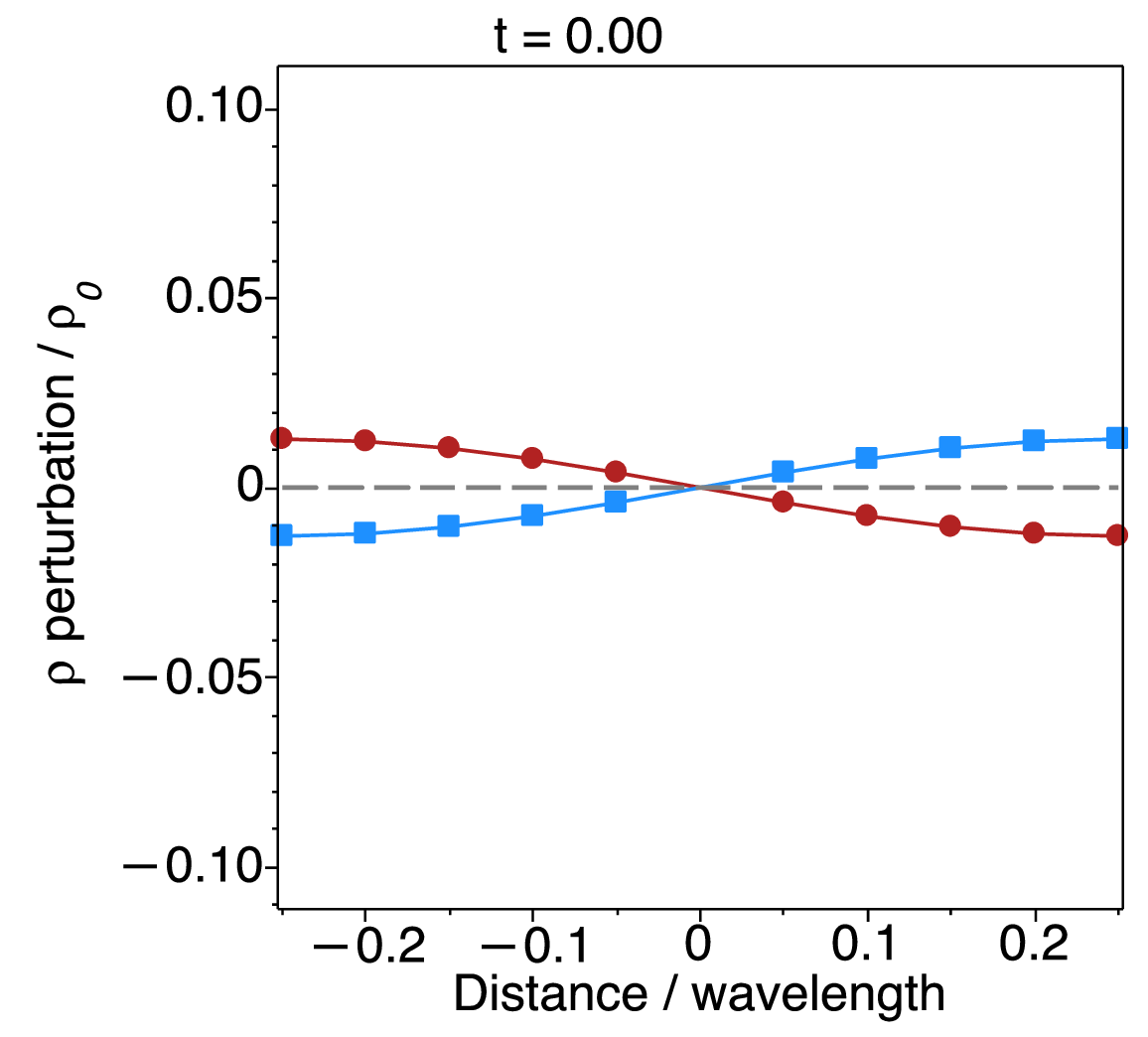}
        \includegraphics[width=0.4\linewidth]{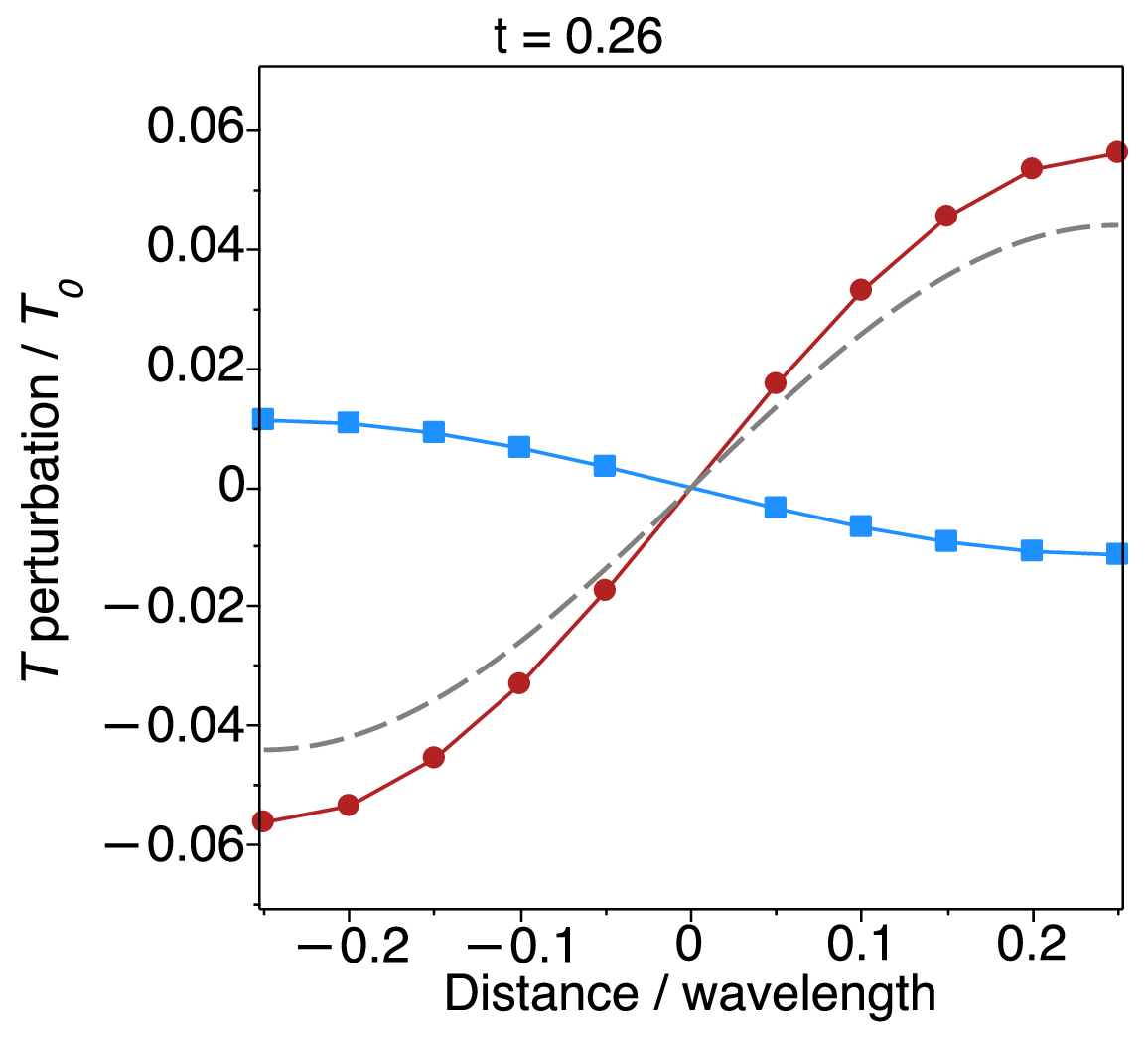}
        \includegraphics[width=0.4\linewidth]{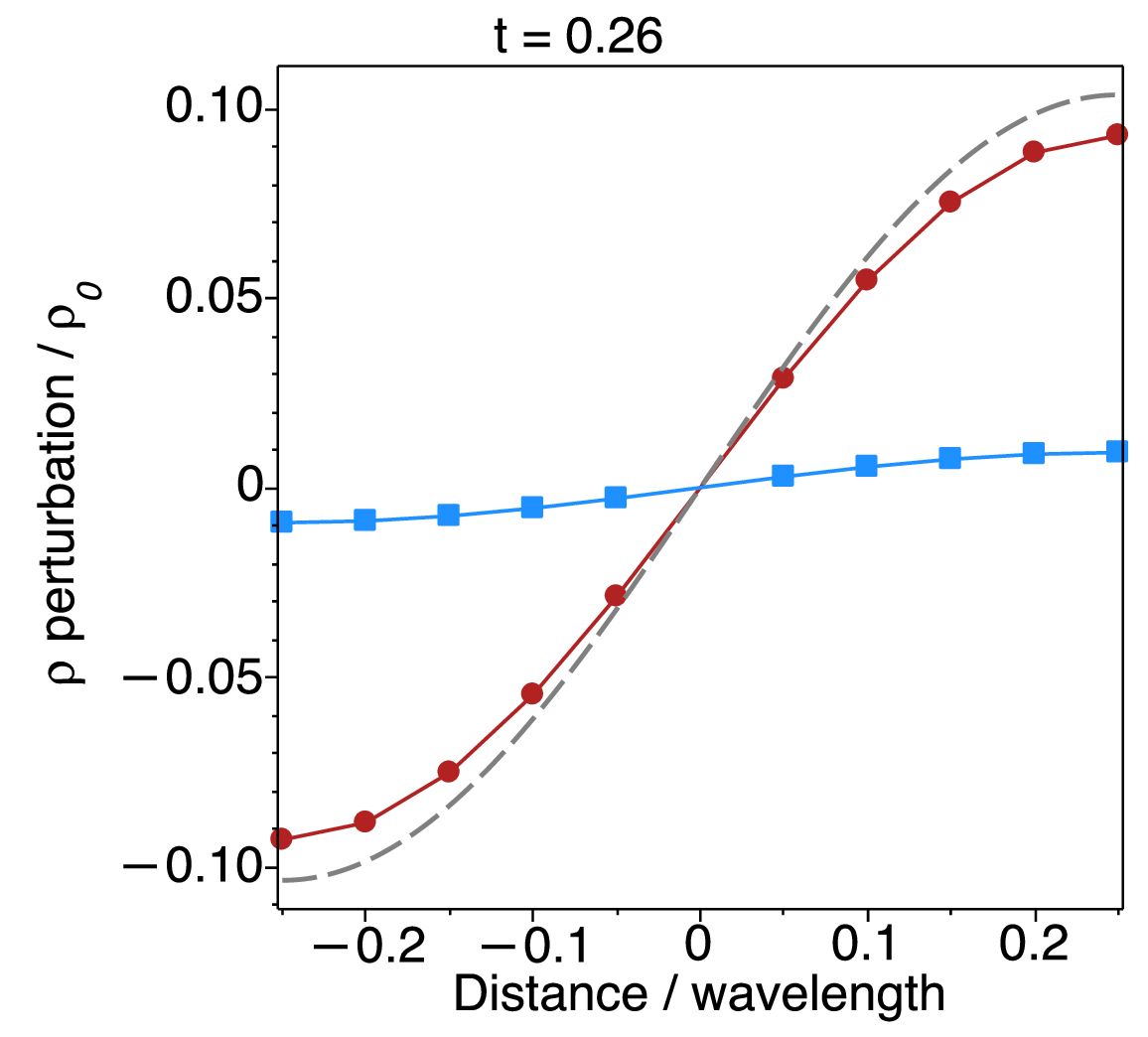}
        
        \includegraphics[width=0.4\linewidth]{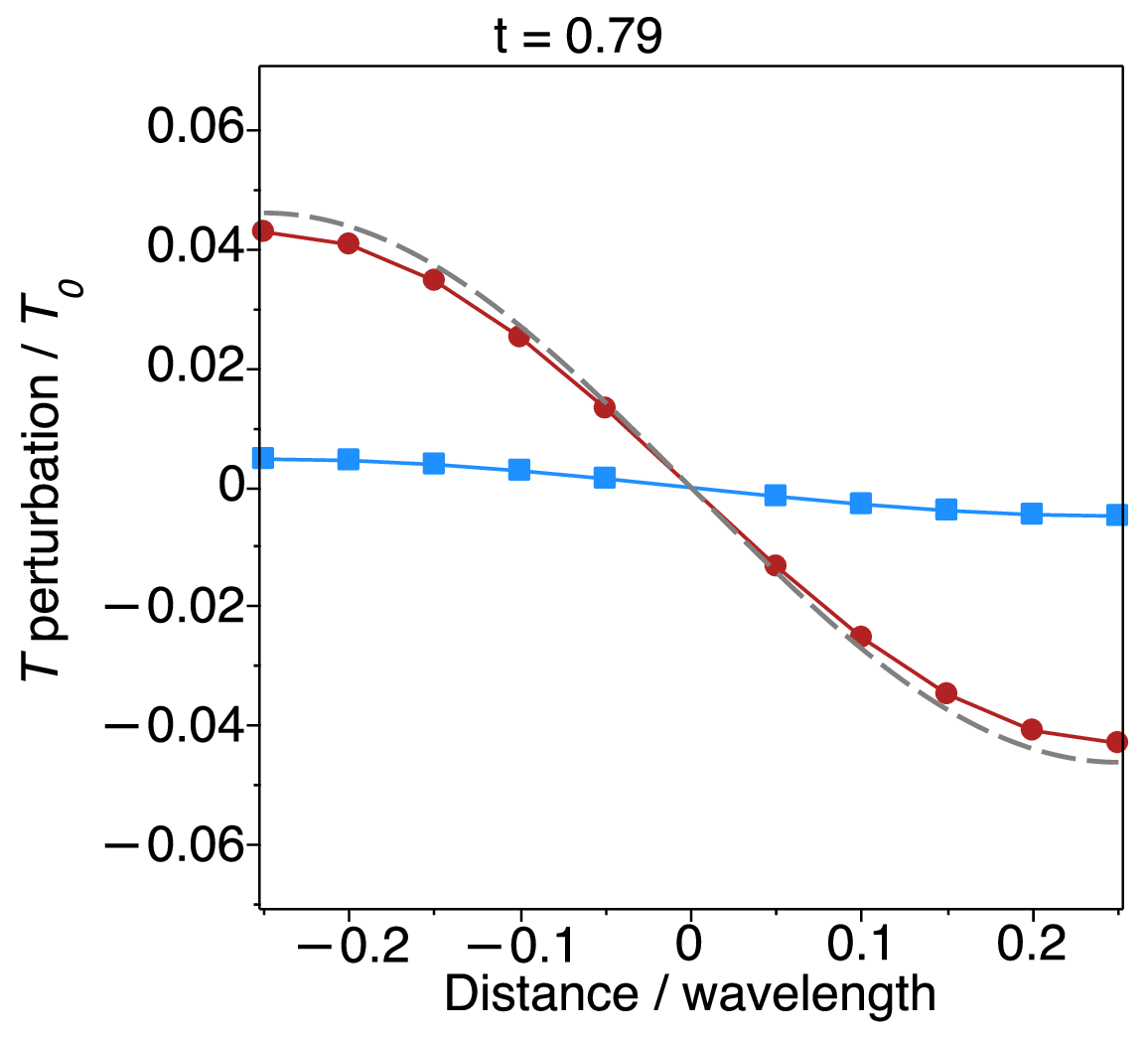}
        \includegraphics[width=0.4\linewidth]{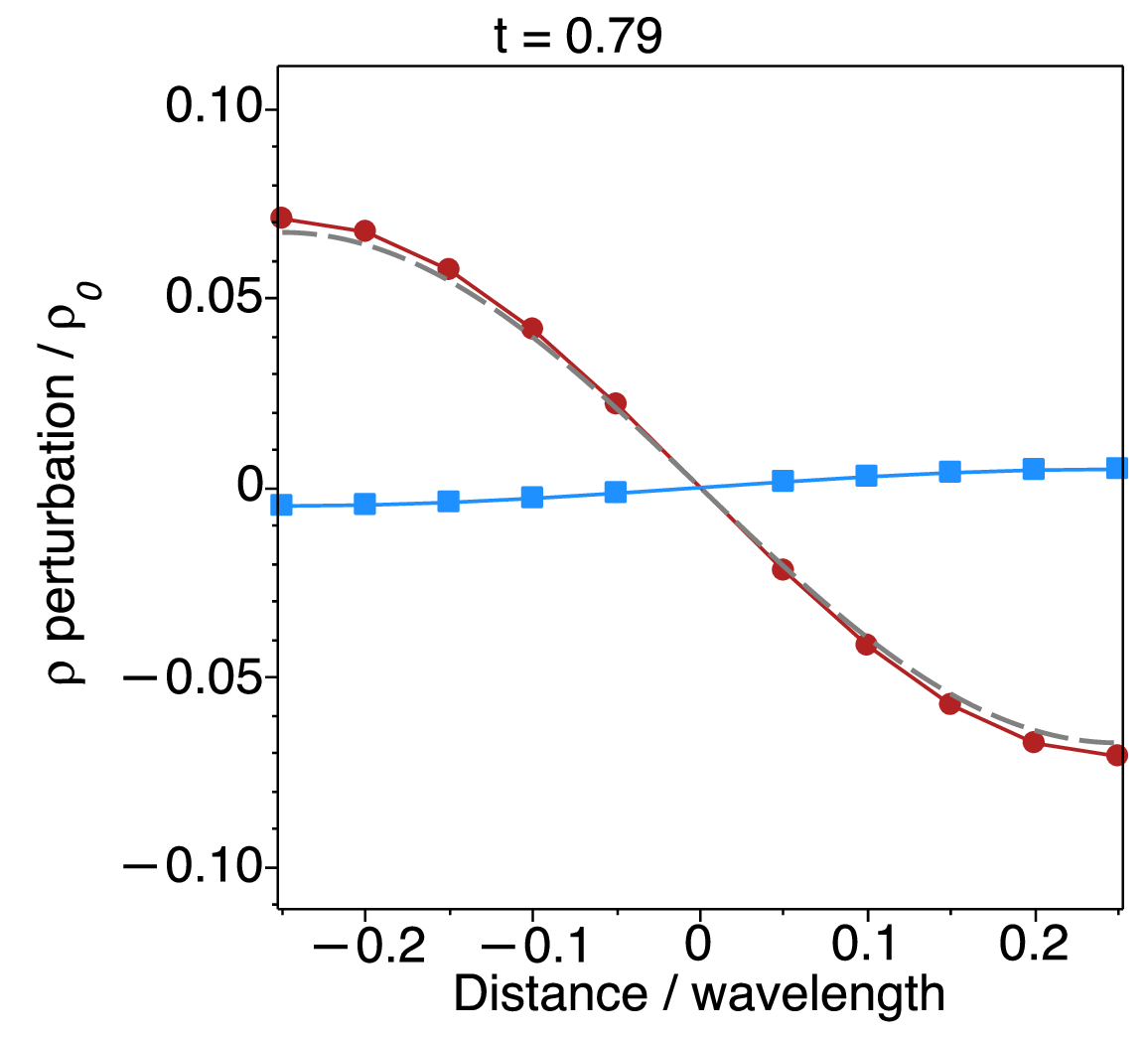}
	\caption{Left: spatio-temporal evolution of the fundamental harmonic of the slow mode (circles, red), entropy mode (boxes, blue), and full numerical solution of Eqs.~(\ref{eq:motion})--(\ref{eq:energy}) (grey dashed) for loop temperature perturbation. The time-dependence of the slow and entropy modes is given by Eqs.~(\ref{eq:entropy_sol})--(\ref{eq:slow_sol}); the spatial structure of their amplitudes $A_{T,\mathrm{slow}}(z)$ and $A_{T,\mathrm{entropy}}(z)$ is obtained as described in Section~\ref{sec:spatial}.
	{Right: same as left column, but for plasma density perturbations.}
    }
	\label{fig:spatial}
\end{figure}

To obtain the spatial structure $A_{T,\mathrm{slow}}(z)$ and $A_{T,\mathrm{entropy}}(z)$ of the slow and entropy modes given by Eqs.~(\ref{eq:entropy_sol})--(\ref{eq:slow_sol}) in the considered standing wave, we can perform the analysis described in Sections~\ref{sec:damping_time} and \ref{sec:shifts} for $z_0$ varying from the left to right boundary of the resonator, [$-0.25$, 0.25] (in units of the wavelength). In other words, using our numerical solution of Eqs.(\ref{eq:motion})--(\ref{eq:energy}), for each $z_0$ along the loop, we extract the time evolution of the total temperature perturbation, best-fit its envelope with a single exponentially decaying function to obtain $A_{T,\mathrm{slow}}(z)$; followed by best-fitting by the refined model given by Eq.~(\ref{eq:slow_ent_model}) which provides $A_{T,\mathrm{entropy}}(z)$. We note that only the initial amplitudes $A_{T,\mathrm{slow}}$ and $A_{T,\mathrm{entropy}}$ are found to vary with $z$, while the corresponding damping times $\tau_\mathrm{slow}$ and $\tau_\mathrm{entropy}$ and the phase shift $\Delta\phi_\mathrm{diss}$ given by Eq.~(\ref{eq:shift_diss}) remain respectively around 1.95 (standard deviation $\sim 10^{-7}$), 0.6 (standard deviation $\sim 10^{-5}$), and 0.26 (standard deviation $\sim 10^{-5}$), i.e. practically constant,  as expected for a standing wave. Thus, substituting the empirically estimated dependencies $A_{T,\mathrm{slow}}(z)$ and $A_{T,\mathrm{entropy}}(z)$ into Eqs.~(\ref{eq:entropy_sol})--(\ref{eq:slow_sol}), we obtain a full spatio-temporal picture of the temperature evolution in the slow and entropy mode, illustrated in Fig.~\ref{fig:spatial} \textbf{(left column)}. Both modes exhibit an antisymmetric structure about $z=0$. The entropy mode undergoes purely exponential decay without phase change, rapidly flattening along the $z$-axis. In contrast, the slow mode shows oscillatory behaviour with alternating phase and comparatively weaker damping, eventually dominating the solution.
{The right column in Fig.~\ref{fig:spatial} shows a similar analysis of the spatio-temporal evolution of the slow and entropy-mode components in plasma density perturbations, adopting Eqs.~(\ref{eq:entropy_sol})--(\ref{eq:shift_diss}) for $A_{\rho,\mathrm{entropy}}$ and $A_{\rho,\mathrm{slow}}$ as described in Section~\ref{sec:shifts}.
}

\begin{figure}
	\centering
       \includegraphics[width=0.465\linewidth]{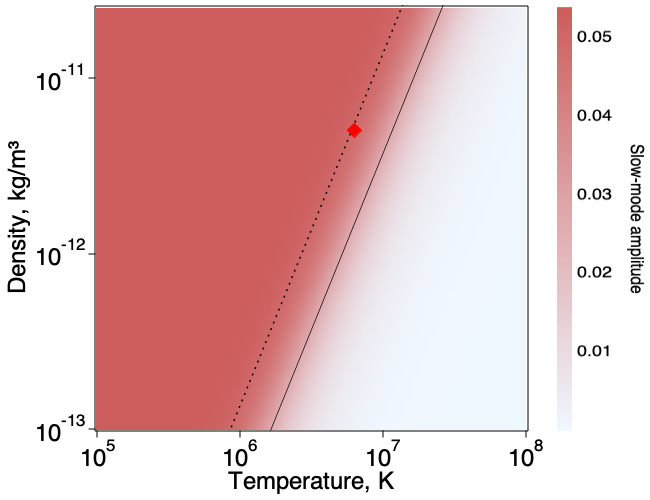}
        \includegraphics[width=0.4\linewidth]{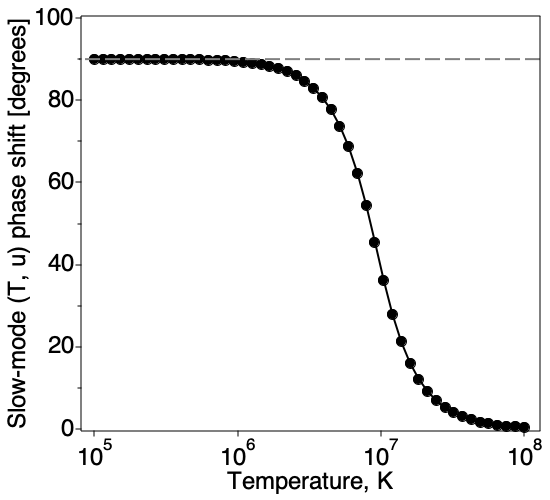}
        \includegraphics[width=0.465\linewidth]{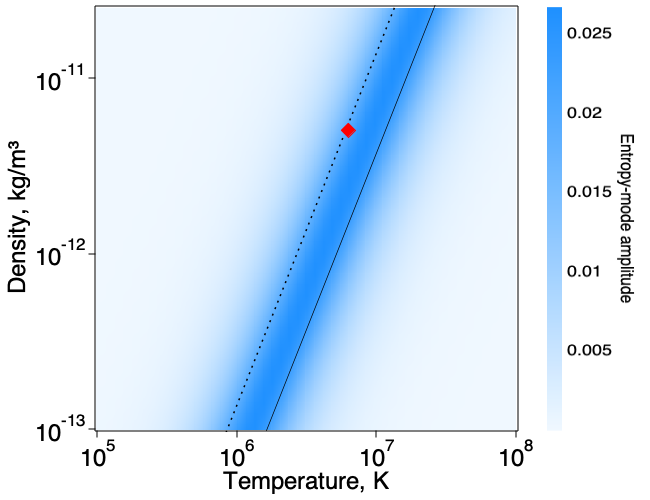}
        \includegraphics[width=0.4\linewidth]{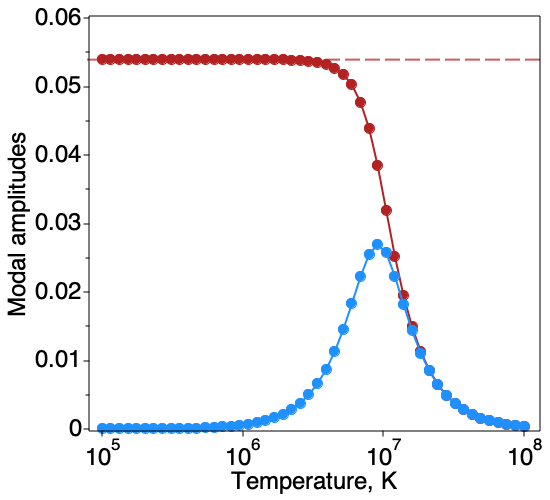}
	\caption{Top: slow-mode temperature amplitude (left) and phase shift between temperature and velocity perturbations (right) given by Eqs.~(\ref{eq:init_amp_anal})--(\ref{eq:diss_shift_anal}). The horizontal dashed line in the top right panel indicates the phase shift of $\pi/2$ in the ideal adiabatic regime.
    Bottom: entropy-mode temperature amplitude as a function of loop temperature and density given by Eq.~(\ref{eq:ent_amp_anal}) (left) and that vs. slow-mode temperature amplitude at fixed density $\rho_0=5\times10^{-12}$\,kg\,m$^{-3}$ (right, in blue and red, respectively). The horizontal dashed line in the bottom right panel shows the adiabatic slow-mode temperature amplitude $(A_{u0}/c_\mathrm{s})\times(\gamma-1)\times\sin(2\pi z_0/\lambda_0)\approx 0.054$ for $A_{u0}/c_\mathrm{s}=0.1$ and $z_0=0.15\lambda_0$.
    The dotted and solid black lines in the left panels indicate the entropy to slow damping time ratio $\tau_\mathrm{entropy}/\tau_\mathrm{slow} = 0.3$ and $0.1$, respectively (see Fig.~\ref{fig:damping_slow_entropy}). The red diamond shows the combination of $\rho_0=5\times10^{-12}$\,kg\,m$^{-3}$ and $T_0=6.3$\,MK used in our numerical analysis (Figs.~\ref{fig:envelopes}--\ref{fig:spatial}).
	}
	\label{fig:amps2d}
\end{figure}

Assuming the standing-wave form $u(z,t)=A_{u0}\cos(kz)\exp(-i\omega t)$ and $T(z,t)=A_{T0}\sin(kz)\exp[-i(\omega t+\Delta\phi_\mathrm{diss})]$ in Eqs.~(\ref{eq:motion})--(\ref{eq:energy}), one can derive an explicit relationship between the real amplitudes of the velocity and temperature perturbations in the slow mode, $A_{u0}$ and $A_{T0}$,
\begin{equation}\label{eq:init_amp_anal}
\frac{A_{T0}}{T_0} = \frac{\gamma-1}{\sqrt{1+(\omega\tau_\mathrm{cond})^{-2}}}\frac{A_{u0}}{c_\mathrm{s}}, \end{equation}
and the conduction-induced phase shift (temperature lagging velocity by $\Delta\phi_\mathrm{diss}$), \begin{equation}\label{eq:diss_shift_anal} \Delta\phi_\mathrm{diss}=\arctan\left(\frac{1}{\omega\tau_\mathrm{cond}}\right).
\end{equation}
The dimensionless parameter $\omega\tau_\mathrm{cond}$ in Eqs.~(\ref{eq:init_amp_anal})--(\ref{eq:diss_shift_anal}) serves as the measure of the thermal conduction effectiveness. Thus, the ratio of the relative amplitudes $(A_{T0}/T_0)/(A_{u0}/c_\mathrm{s})$ may vary from $\gamma-1$ for weak thermal conduction \citep[$\omega\tau_\mathrm{cond} \gg 1$, see e.g. Eq.~(18) in][where non-adiabatic effects are neglected]{2017ApJ...849...62N} to zero for strong conduction ($\omega\tau_\mathrm{cond} \ll 1$, i.e. the isothermal regime with strongly suppressed temperature perturbations). Likewise, the dissipative phase shift $\Delta\phi_\mathrm{diss}$ increases from zero to $\pi/2$ as thermal conduction becomes more effective \citep[decreasing $\omega\tau_\mathrm{cond}$, see e.g.][]{2023FrASS..1067781Z}. The top row panels in Fig.~\ref{fig:amps2d} show the slow-mode amplitude $A_{T,\mathrm{slow}}=(A_{T0}/T_0)\sin(2\pi z_0/\lambda_0)$ at $z_0=0.15\lambda_0$, obtained from Eq.~(\ref{eq:init_amp_anal}) for $A_{u0}/c_\mathrm{s} = 0.1$, and the total phase shift $\Delta\phi_{u-T}=\Delta\phi_\mathrm{ideal} - \Delta\phi_\mathrm{diss}$ between plasma temperature and velocity perturbations in the considered standing wave ($\Delta\phi_\mathrm{ideal}=\pi/2$), obtained from Eq.~(\ref{eq:diss_shift_anal}). Thus, both $A_{T,\mathrm{slow}}$ and $\Delta\phi_{u-T}$ are seen to vary from their adiabatic values $0.1\times(\gamma-1)\times\sin(2\pi\times0.15)\approx 0.054$ and $90^\circ$, respectively, to zero in the strongly conductive (tending to isothermal) limit.

Using the link between $\Delta\phi_\mathrm{diss}$ and the amplitude of the entropy mode $A_{T,\mathrm{entropy}}$ established by Eq.~(\ref{eq:shift_diss}) and the analytical solutions given by Eqs.~(\ref{eq:init_amp_anal})--(\ref{eq:diss_shift_anal}), we can obtain $A_{T,\mathrm{entropy}}$ as a function of the dimensionless parameter $\omega\tau_\mathrm{cond}$ (prescribed by the equilibrium plasma temperature $T_0$, density $\rho_0$, and loop length $L$) and the initial perturbation amplitude $A_{u0}/c_\mathrm{s}$,
\begin{equation}\label{eq:ent_amp_anal}
    A_{T,\mathrm{entropy}} = A_{T,\mathrm{slow}}\sin\left(\arctan\left[\frac{1}{\omega\tau_\mathrm{cond}}\right]\right).
\end{equation}
Bottom left panel in Fig.~\ref{fig:amps2d} shows $A_{T,\mathrm{entropy}}$ given by Eq.~(\ref{eq:ent_amp_anal}) at $z_0=0.15\lambda_0$, for a broad range of the equilibrium plasma temperature $T_0$ and density $\rho_0$, loop length $L=180$\,Mm and initial velocity perturbation amplitude $A_{u0}/c_\mathrm{s} = 0.1$. Bottom right panel shows the comparison of the modal amplitudes, $A_{T,\mathrm{entropy}}$ and $A_{T,\mathrm{slow}}$ across plasma temperature $T_0$ (fixed $\rho_0$). We observe a well-defined region in the ($T_0$, $\rho_0$) parameter space, spanning approximately one order of magnitude in temperature, where the entropy mode is excited most efficiently.

\section{Conclusions} \label{sec:concl}
We studied the time-domain evolution of compressive oscillations in a coronal loop in the presence of thermal conduction, with particular emphasis on the simultaneous development of slow magnetoacoustic and entropy modes. Our analysis provides a proof of concept that the intrinsically non-oscillatory entropy mode leaves {potentially} identifiable signatures in the temporal behaviour of compressive perturbations. In particular, we demonstrate that these signatures manifest through a non-exponential damping and the envelope asymmetry of the total temperature {{and density}} perturbations, and phase shifts between the associated slow-mode temperature{{/density}} and velocity oscillations, thereby making the entropy mode potentially accessible to observational analysis. The main results can be summarised as follows.

\begin{itemize}
    \item From dispersion analysis, we derived damping times of both slow and entropy modes in the coronal plasma with thermal conduction, see $\tau_\mathrm{slow}$ and $\tau_\mathrm{entropy}$ given by Eqs.~(\ref{eq:dampt-slow})--(\ref{eq:dampt-ent}). The damping time of the entropy mode is shown to be always shorter (by at least three times) than that of the slow mode (Fig.~\ref{fig:damping_slow_entropy}). 
    
    \item Simultaneous development of two exponentially decaying modes makes the total temperature {{and density}} perturbation envelopes essentially non-exponential (Fig.~\ref{fig:envelopes}), which can serve as the characteristic signature of the entropy mode in observations.
    An apparent switch time between non-exponential and exponential damping regimes is shown to coincide with $3\tau_\mathrm{entropy}$ given by Eq.~(\ref{eq:dampt-ent}).
    A revision of existing catalogues of SUMER-type oscillations on the Sun \citep[e.g.][]{2019ApJ...874L...1N} and other stars \citep[in the form of quasi-periodic pulsations in flares, see e.g.][]{2016ApJ...830..110C} in this context seems of interest.
    {By analogy with the switch time between Gaussian and exponential damping profiles of kink oscillations \citep[e.g.][]{2017A&A...600A..78P}, the use of Bayesian analysis with Markov chain Monte Carlo sampling method \citep{2018AdSpR..61..655A, 2022SSRv..218....9A} seems especially relevant for the identification of the damping regime switch time by the entropy mode and its discrimination from other possible mechanisms.}
    We note that the previously proposed scenarios of apparent damping of propagating slow waves via line-of-sight integration and loop's cross-field multi-thermality \citep{2024A&A...683A.109V, 2024ApJ...970...58K, 2024SoPh..299....2F} cannot explain the non-exponential damping of standing compressive waves.
    {The development of nonlinear effects in standing compressive oscillations \citep[e.g.][]{2008ApJ...685.1286V, 2013A&A...553A..23R} is shown to be effectively suppressed by viscosity \citep{2018ApJ...860..107W}, hence is not likely to affect the damping pattern and obscure the entropy-mode signatures identified in this work.}
    We showed two-step best-fitting procedure for estimating the damping times of both the slow and entropy modes from the analysis of temperature perturbation envelopes.
    
    \item Due to a short-lived nature of the entropy mode, the uncertainty of its damping time estimation in our numerical analysis is found to be almost twelve times larger than that of the slow mode ($\tau^\mathrm{fit}_\mathrm{slow}=1.95\pm 0.002$ vs. $\tau^\mathrm{fit}_\mathrm{entropy}=0.6\pm 0.024$ for a loop with $\rho_0=5\times10^{-12}$\,kg\,m$^{-3}$, $T_0=6.3$\,MK, and $L=180$\,Mm, both are normalised to $P_0=15.7$\,min).
    {If we neglect the effect of the entropy mode and best-fit the entire envelope (not only at $t>3\tau_\mathrm{entropy}$) by the exponentially decaying function (as it is usually used in observations), we obtain $\tau^\mathrm{fit}_\mathrm{slow}=1.96\pm 0.008$, providing the substantial increase (by a factor of four) in the slow-mode damping time uncertainty \citep[cf.][where a similar problem was acknowledged for kink oscillations due to neglecting Gaussian damping phase]{Nechaeva2019Catalog}.}
    This is likely to worsen in real observational conditions, e.g. due to noise, finite cadence, line-of-sight integration effects, etc.
    {For example, using $\tau_\mathrm{slow}/\tau_\mathrm{entropy} = 3$ and taking the observed standing slow-wave damping time of $1598 \pm 138$\,s \citep{2021ApJ...914...81K} and correcting the detected uncertainty by a factor of four to 35\,s and propagating it to the potential entropy damping time estimate ($12 \times 35$\,s, according to our analysis), we obtain $\tau_\mathrm{entropy}=532\pm420$\,s.}
    On the other hand, we demonstrated that the entropy mode contributes additively to the overall temperature perturbation envelope, i.e. it enhances the lower envelope and reduces the upper envelope (by the same amount), thereby making them essentially asymmetric (see Fig.~\ref{fig:envelopes}). Hence, taking the mean of the total signal's upper and lower envelopes seems another potentially promising method to identify the entropy mode in observations, which needs to be tested observationally. 
    
    \item Despite its rapid decay, the entropy mode affects the subsequent evolution of the slow mode, accelerating the development of the temperature perturbation from the initial velocity perturbation, which is seen as a reduction of the phase shift between them from an adiabatic value of $\pi/2$ (see Eq.~(\ref{eq:shift_diss}) and Fig.~\ref{fig:shifts}). This makes it more feasible to capture the entropy-mode effect in observations, even if the initial signal evolution does not allow for a meaningful detection \citep[e.g. due to impulsive excitation and transition processes,][]{2005A&A...436..701S, 2019A&A...628A.133K}.
    {Detecting a correlation between the signal's envelope asymmetry (quantified via e.g. $A_\mathrm{entropy}$ and $A_\mathrm{slow}$) and phase shifts, as prescribed by Eq.~(\ref{eq:shift_diss}), would provide strong evidence of the entropy-mode effect in observations.}
    {We also considered the combined entropy-mode effect of plasma density and temperature perturbations on the synthetic optically-thin intensity, which is found to be wavelength-dependent. In particular, for the considered model parameters, it enhances in the 193\,\AA\ channel and suppresses in the 94\,\AA\ of SDO/AIA.}
    Extending this work towards the manifestation of the entropy mode in other plasma parameters, such as pressure and potentially magnetic field \citep[e.g. within the thin flux tube model, see e.g.][]{2021A&A...646A.155D} is {also instructive}.
    Likewise, Eq.~(\ref{eq:shift_diss}) is specifically derived for flow-induced perturbations (with $u(z,0)\ne0$ and $T(z,0)=0$). Its generalisation for other excitation mechanisms of standing compressive oscillations is also of interest. {In particular, the ratio of the entropy to slow wave magnitudes was previously shown to be sensitive to the type of the initial perturbation \citep[see][Eq.~(22) and Fig.~2]{2023FrASS..1067781Z}. More work is required to understand how this propagates into our Eq.~(\ref{eq:shift_diss}).}
    Such an association of the entropy mode with phase shifts in compressive waves allows us to suggest that the non-adiabatic mechanisms that do not affect the wave phase \citep[such as viscosity, see][]{2019ApJ...886....2W} are not likely to induce the entropy mode in a loop, but this suggestion requires detailed analysis.
    
    \item According to Fig.~\ref{fig:amps2d}, there is a well-defined region in the ($T_0$, $\rho_0$) parametric space where the entropy mode is effectively excited and can therefore be potentially detected. But, one needs to account for its damping rate. Our examples in Figs.~\ref{fig:envelopes}--\ref{fig:spatial} are produced for the case when the ratio of damping times $\tau_\mathrm{entropy}/\tau_\mathrm{slow}$ remains about 0.3 and the entropy-mode amplitude $A_{T,\mathrm{entropy}}\approx 0.014$. Going towards higher entropy mode amplitudes in the ($T_0$, $\rho_0$) space (the maximum $A_{T,\mathrm{entropy}}\approx 0.03$ as seen in Fig.~\ref{fig:amps2d}) may result in shortening of its damping time and hence, more problematic detection. We also note that this picture of the entropy mode amplitude dependence on the loop's equilibrium state may readily change by the inclusion of other coronal heating and cooling processes and the wave-induced misbalance between them{, e.g. in denser loops with $\rho_0 \gtrsim 10^{-11}$\,kg\,m$^{-3}$} \citep[][]{2025LRSP...22....4K, 2024mpsp.book..415A, 2021PPCF...63l4008K}.
\end{itemize}

These results suggest that the entropy mode is not merely a theoretical construct, but a physically relevant and {potentially} observable component of coronal dynamics. Revisiting existing observations, together with future high-cadence and high-sensitivity data and refined forward modelling, will be crucial for systematically testing these signatures and incorporating the entropy mode into the broader framework of MHD coronal seismology.


\begin{acknowledgments}
The work is supported by the UKRI Stephen
Hawking Fellowship EP/Z535473/1 and the Latvian Council of Science Project lzp-2024/1-0023 (DK), the STFC Consolidated Grant ST/X000915/1 (SB), and STFC Studentship Grant UKRI1788 (MS). DK and MS also thank the Undergraduate Research Support Scheme from the University of Warwick.
For the purpose of open access, the authors have applied a Creative Commons Attribution (CC BY) license to any author-accepted manuscript version arising.
\end{acknowledgments}

\bibliographystyle{aasjournal}

\begin{thebibliography}{}
	\expandafter\ifx\csname natexlab\endcsname\relax\def\natexlab#1{#1}\fi
	\providecommand{\url}[1]{\href{#1}{#1}}
	\providecommand{\dodoi}[1]{doi:~\href{http://doi.org/#1}{\nolinkurl{#1}}}
	\providecommand{\doeprint}[1]{\href{http://ascl.net/#1}{\nolinkurl{http://ascl.net/#1}}}
	\providecommand{\doarXiv}[1]{\href{https://arxiv.org/abs/#1}{\nolinkurl{https://arxiv.org/abs/#1}}}
	
	\bibitem[{{Andries} {et~al.}(2009){Andries}, {van Doorsselaere}, {Roberts},
		{Verth}, {Verwichte}, \& {Erd{\'e}lyi}}]{2009SSRv..149....3A}
	{Andries}, J., {van Doorsselaere}, T., {Roberts}, B., {et~al.} 2009, \ssr, 149,
	3, \dodoi{10.1007/s11214-009-9561-2}
	
	\bibitem[{{Anfinogentov} {et~al.}(2022){Anfinogentov}, {Antolin}, {Inglis},
		{Kolotkov}, {Kupriyanova}, {McLaughlin}, {Nistic{\`o}}, {Pascoe}, {Krishna
			Prasad}, \& {Yuan}}]{2022SSRv..218....9A}
	{Anfinogentov}, S.~A., {Antolin}, P., {Inglis}, A.~R., {et~al.} 2022, \ssr,
	218, 9, \dodoi{10.1007/s11214-021-00869-w}
	
	\bibitem[{{Antolin} \& {Froment}(2022)}]{2022FrASS...920116A}
	{Antolin}, P., \& {Froment}, C. 2022, Frontiers in Astronomy and Space
	Sciences, 9, 820116, \dodoi{10.3389/fspas.2022.820116}
	
	\bibitem[{{Arber} {et~al.}(2023){Arber}, {Goffrey}, \&
		{Ridgers}}]{2023FrASS..1055124A}
	{Arber}, T.~D., {Goffrey}, T., \& {Ridgers}, C. 2023, Frontiers in Astronomy
	and Space Sciences, 10, 1155124, \dodoi{10.3389/fspas.2023.1155124}
	
	\bibitem[{{Arregui}(2018)}]{2018AdSpR..61..655A}
	{Arregui}, I. 2018, Advances in Space Research, 61, 655,
	\dodoi{10.1016/j.asr.2017.09.031}
	
	\bibitem[{{Arregui} {et~al.}(2023){Arregui}, {Kolotkov}, \&
		{Nakariakov}}]{2023A&A...677A..23A}
	{Arregui}, I., {Kolotkov}, D.~Y., \& {Nakariakov}, V.~M. 2023, \aap, 677, A23,
	\dodoi{10.1051/0004-6361/202346834}
	
	\bibitem[{{Arregui} \& {Van Doorsselaere}(2024)}]{2024mpsp.book..415A}
	{Arregui}, I., \& {Van Doorsselaere}, T. 2024, in Magnetohydrodynamic Processes
	in Solar Plasmas, ed. A.~K. {Srivastava}, M.~{Goossens}, \& I.~{Arregui},
	415--450, \dodoi{10.1016/B978-0-32-395664-2.00015-3}
	
	\bibitem[{{Aschwanden} {et~al.}(2002){Aschwanden}, {De Pontieu}, {Schrijver},
		\& {Title}}]{2002SoPh..206...99A}
	{Aschwanden}, M.~J., {De Pontieu}, B., {Schrijver}, C.~J., \& {Title}, A.~M.
	2002, \solphys, 206, 99, \dodoi{10.1023/A:1014916701283}
	
	\bibitem[{{Banerjee} {et~al.}(2011){Banerjee}, {Gupta}, \&
		{Teriaca}}]{2011SSRv..158..267B}
	{Banerjee}, D., {Gupta}, G.~R., \& {Teriaca}, L. 2011, \ssr, 158, 267,
	\dodoi{10.1007/s11214-010-9698-z}
	
	\bibitem[{{Banerjee} \& {Krishna Prasad}(2016)}]{2016GMS...216..419B}
	{Banerjee}, D., \& {Krishna Prasad}, S. 2016, Geophysical Monograph Series,
	216, 419, \dodoi{10.1002/9781119055006.ch24}
	
	\bibitem[{{Banerjee} {et~al.}(2021){Banerjee}, {Krishna Prasad}, {Pant},
		{McLaughlin}, {Antolin}, {Magyar}, {Ofman}, {Tian}, {Van Doorsselaere}, {De
			Moortel}, \& {Wang}}]{2021SSRv..217...76B}
	{Banerjee}, D., {Krishna Prasad}, S., {Pant}, V., {et~al.} 2021, \ssr, 217, 76,
	\dodoi{10.1007/s11214-021-00849-0}
	
	\bibitem[{{Belov} {et~al.}(2025){Belov}, {Goffrey}, {Arber}, \&
		{Kolotkov}}]{2025A&A...693A.186B}
	{Belov}, S.~A., {Goffrey}, T., {Arber}, T.~D., \& {Kolotkov}, D.~Y. 2025, \aap,
	693, A186, \dodoi{10.1051/0004-6361/202452938}
	
	\bibitem[{{Belov} {et~al.}(2021){Belov}, {Molevich}, \&
		{Zavershinskii}}]{2021SoPh..296..122B}
	{Belov}, S.~A., {Molevich}, N.~E., \& {Zavershinskii}, D.~I. 2021, \solphys,
	296, 122, \dodoi{10.1007/s11207-021-01868-4}
	
	\bibitem[{{Boerner} {et~al.}(2012){Boerner}, {Edwards}, {Lemen}, {Rausch},
		{Schrijver}, {Shine}, {Shing}, {Stern}, {Tarbell}, {Title}, {Wolfson},
		{Soufli}, {Spiller}, {Gullikson}, {McKenzie}, {Windt}, {Golub}, {Podgorski},
		{Testa}, \& {Weber}}]{2012SoPh..275...41B}
	{Boerner}, P., {Edwards}, C., {Lemen}, J., {et~al.} 2012, \solphys, 275, 41,
	\dodoi{10.1007/s11207-011-9804-8}
	
	\bibitem[{{Boerner} {et~al.}(2014){Boerner}, {Testa}, {Warren}, {Weber}, \&
		{Schrijver}}]{2014SoPh..289.2377B}
	{Boerner}, P.~F., {Testa}, P., {Warren}, H., {Weber}, M.~A., \& {Schrijver},
	C.~J. 2014, \solphys, 289, 2377, \dodoi{10.1007/s11207-013-0452-z}
	
	\bibitem[{{Cho} {et~al.}(2016){Cho}, {Cho}, {Nakariakov}, {Kim}, \&
		{Kumar}}]{2016ApJ...830..110C}
	{Cho}, I.-H., {Cho}, K.-S., {Nakariakov}, V.~M., {Kim}, S., \& {Kumar}, P.
	2016, \apj, 830, 110, \dodoi{10.3847/0004-637X/830/2/110}
	
	\bibitem[{{De Moortel}(2009)}]{2009SSRv..149...65D}
	{De Moortel}, I. 2009, \ssr, 149, 65, \dodoi{10.1007/s11214-009-9526-5}
	
	\bibitem[{{De Moortel} \& {Hood}(2003)}]{2003A&A...408..755D}
	{De Moortel}, I., \& {Hood}, A.~W. 2003, \aap, 408, 755,
	\dodoi{10.1051/0004-6361:20030984}
	
	\bibitem[{{Dolla} {et~al.}(2012){Dolla}, {Marqu{\'e}}, {Seaton}, {Van
			Doorsselaere}, {Dominique}, {Berghmans}, {Cabanas}, {De Groof}, {Schmutz},
		{Verdini}, {West}, {Zender}, \& {Zhukov}}]{2012ApJ...749L..16D}
	{Dolla}, L., {Marqu{\'e}}, C., {Seaton}, D.~B., {et~al.} 2012, \apjl, 749, L16,
	\dodoi{10.1088/2041-8205/749/1/L16}
	
	\bibitem[{{Duckenfield} {et~al.}(2018){Duckenfield}, {Anfinogentov}, {Pascoe},
		\& {Nakariakov}}]{2018ApJ...854L...5D}
	{Duckenfield}, T., {Anfinogentov}, S.~A., {Pascoe}, D.~J., \& {Nakariakov},
	V.~M. 2018, \apjl, 854, L5, \dodoi{10.3847/2041-8213/aaaaeb}
	
	\bibitem[{{Duckenfield} {et~al.}(2019){Duckenfield}, {Goddard}, {Pascoe}, \&
		{Nakariakov}}]{2019A&A...632A..64D}
	{Duckenfield}, T.~J., {Goddard}, C.~R., {Pascoe}, D.~J., \& {Nakariakov}, V.~M.
	2019, \aap, 632, A64, \dodoi{10.1051/0004-6361/201936822}
	
	\bibitem[{{Duckenfield} {et~al.}(2021){Duckenfield}, {Kolotkov}, \&
		{Nakariakov}}]{2021A&A...646A.155D}
	{Duckenfield}, T.~J., {Kolotkov}, D.~Y., \& {Nakariakov}, V.~M. 2021, \aap,
	646, A155, \dodoi{10.1051/0004-6361/202039791}
	
	\bibitem[{{Fang} {et~al.}(2015){Fang}, {Yuan}, {Van Doorsselaere}, {Keppens},
		\& {Xia}}]{2015ApJ...813...33F}
	{Fang}, X., {Yuan}, D., {Van Doorsselaere}, T., {Keppens}, R., \& {Xia}, C.
	2015, \apj, 813, 33, \dodoi{10.1088/0004-637X/813/1/33}
	
	\bibitem[{{Fedenev} {et~al.}(2024){Fedenev}, {Nakariakov}, \&
		{Anfinogentov}}]{2024SoPh..299....2F}
	{Fedenev}, V.~V., {Nakariakov}, V.~M., \& {Anfinogentov}, S.~A. 2024, \solphys,
	299, 2, \dodoi{10.1007/s11207-023-02246-y}
	
	\bibitem[{{Goddard} \& {Nakariakov}(2016)}]{2016A&A...590L...5G}
	{Goddard}, C.~R., \& {Nakariakov}, V.~M. 2016, \aap, 590, L5,
	\dodoi{10.1051/0004-6361/201628718}
	
	\bibitem[{{Goossens} {et~al.}(2006){Goossens}, {Andries}, \&
		{Arregui}}]{2006RSPTA.364..433G}
	{Goossens}, M., {Andries}, J., \& {Arregui}, I. 2006, Philosophical
	Transactions of the Royal Society of London Series A, 364, 433,
	\dodoi{10.1098/rsta.2005.1708}
	
	\bibitem[{{Goossens} {et~al.}(2002){Goossens}, {Andries}, \&
		{Aschwanden}}]{2002A&A...394L..39G}
	{Goossens}, M., {Andries}, J., \& {Aschwanden}, M.~J. 2002, \aap, 394, L39,
	\dodoi{10.1051/0004-6361:20021378}
	
	\bibitem[{{Hood} {et~al.}(2013){Hood}, {Ruderman}, {Pascoe}, {De Moortel},
		{Terradas}, \& {Wright}}]{2013A&A...551A..39H}
	{Hood}, A.~W., {Ruderman}, M., {Pascoe}, D.~J., {et~al.} 2013, \aap, 551, A39,
	\dodoi{10.1051/0004-6361/201220617}
	
	\bibitem[{{Inglis} \& {Nakariakov}(2009)}]{2009A&A...493..259I}
	{Inglis}, A.~R., \& {Nakariakov}, V.~M. 2009, \aap, 493, 259,
	\dodoi{10.1051/0004-6361:200810473}
	
	\bibitem[{{Keppens} {et~al.}(2025){Keppens}, {Zhou}, \&
		{Xia}}]{2025LRSP...22....4K}
	{Keppens}, R., {Zhou}, Y., \& {Xia}, C. 2025, Living Reviews in Solar Physics,
	22, 4, \dodoi{10.1007/s41116-025-00043-2}
	
	\bibitem[{{Kolotkov}(2022)}]{2022FrASS...973664K}
	{Kolotkov}, D.~Y. 2022, Frontiers in Astronomy and Space Sciences, 9, 402,
	\dodoi{10.3389/fspas.2022.1073664}
	
	\bibitem[{{Kolotkov} \& {Nakariakov}(2022)}]{2022MNRAS.514L..51K}
	{Kolotkov}, D.~Y., \& {Nakariakov}, V.~M. 2022, \mnras, 514, L51,
	\dodoi{10.1093/mnrasl/slac054}
	
	\bibitem[{{Kolotkov} {et~al.}(2023){Kolotkov}, {Nakariakov}, \&
		{Fihosy}}]{2023Physi...5..193K}
	{Kolotkov}, D.~Y., {Nakariakov}, V.~M., \& {Fihosy}, J.~B. 2023, Physics, 5,
	193, \dodoi{10.3390/physics5010015}
	
	\bibitem[{{Kolotkov} {et~al.}(2021{\natexlab{a}}){Kolotkov}, {Nakariakov},
		{Moss}, \& {Shellard}}]{2021MNRAS.505.3505K}
	{Kolotkov}, D.~Y., {Nakariakov}, V.~M., {Moss}, G., \& {Shellard}, P.
	2021{\natexlab{a}}, \mnras, 505, 3505, \dodoi{10.1093/mnras/stab1587}
	
	\bibitem[{{Kolotkov} {et~al.}(2019){Kolotkov}, {Nakariakov}, \&
		{Zavershinskii}}]{2019A&A...628A.133K}
	{Kolotkov}, D.~Y., {Nakariakov}, V.~M., \& {Zavershinskii}, D.~I. 2019, \aap,
	628, A133, \dodoi{10.1051/0004-6361/201936072}
	
	\bibitem[{{Kolotkov} {et~al.}(2021{\natexlab{b}}){Kolotkov}, {Zavershinskii},
		\& {Nakariakov}}]{2021PPCF...63l4008K}
	{Kolotkov}, D.~Y., {Zavershinskii}, D.~I., \& {Nakariakov}, V.~M.
	2021{\natexlab{b}}, Plasma Physics and Controlled Fusion, 63, 124008,
	\dodoi{10.1088/1361-6587/ac36a5}
	
	\bibitem[{{Krishna Prasad} {et~al.}(2014){Krishna Prasad}, {Banerjee}, \& {Van
			Doorsselaere}}]{2014ApJ...789..118K}
	{Krishna Prasad}, S., {Banerjee}, D., \& {Van Doorsselaere}, T. 2014, \apj,
	789, 118, \dodoi{10.1088/0004-637X/789/2/118}
	
	\bibitem[{{Krishna Prasad} \& {Van Doorsselaere}(2021)}]{2021ApJ...914...81K}
	{Krishna Prasad}, S., \& {Van Doorsselaere}, T. 2021, \apj, 914, 81,
	\dodoi{10.3847/1538-4357/abfb01}
	
	\bibitem[{{Krishna Prasad} \& {Van Doorsselaere}(2024)}]{2024ApJ...970...58K}
	---. 2024, \apj, 970, 58, \dodoi{10.3847/1538-4357/ad54b7}
	
	\bibitem[{{Kumar} {et~al.}(2013){Kumar}, {Innes}, \&
		{Inhester}}]{2013ApJ...779L...7K}
	{Kumar}, P., {Innes}, D.~E., \& {Inhester}, B. 2013, \apjl, 779, L7,
	\dodoi{10.1088/2041-8205/779/1/L7}
	
	\bibitem[{{Kumar} {et~al.}(2015){Kumar}, {Nakariakov}, \&
		{Cho}}]{2015ApJ...804....4K}
	{Kumar}, P., {Nakariakov}, V.~M., \& {Cho}, K.-S. 2015, \apj, 804, 4,
	\dodoi{10.1088/0004-637X/804/1/4}
	
	\bibitem[{{Kupriyanova} {et~al.}(2019){Kupriyanova}, {Kashapova}, {Van
			Doorsselaere}, {Chowdhury}, {Srivastava}, \& {Moon}}]{2019MNRAS.483.5499K}
	{Kupriyanova}, E.~G., {Kashapova}, L.~K., {Van Doorsselaere}, T., {et~al.}
	2019, \mnras, 483, 5499, \dodoi{10.1093/mnras/sty3480}
	
	\bibitem[{{Li} {et~al.}(2020){Li}, {Antolin}, {Guo}, {Kuznetsov}, {Pascoe},
		{Van Doorsselaere}, \& {Vasheghani Farahani}}]{2020SSRv..216..136L}
	{Li}, B., {Antolin}, P., {Guo}, M.-Z., {et~al.} 2020, \ssr, 216, 136,
	\dodoi{10.1007/s11214-020-00761-z}
	
	\bibitem[{{Li} {et~al.}(2017){Li}, {Liu}, \& {Vai Tam}}]{2017ApJ...842...99L}
	{Li}, H., {Liu}, Y., \& {Vai Tam}, K. 2017, \apj, 842, 99,
	\dodoi{10.3847/1538-4357/aa7677}
	
	\bibitem[{{Magyar} \& {Van Doorsselaere}(2016)}]{2016A&A...595A..81M}
	{Magyar}, N., \& {Van Doorsselaere}, T. 2016, \aap, 595, A81,
	\dodoi{10.1051/0004-6361/201629010}
	
	\bibitem[{{Mandal} {et~al.}(2018){Mandal}, {Krishna Prasad}, \&
		{Banerjee}}]{2018ApJ...853..134M}
	{Mandal}, S., {Krishna Prasad}, S., \& {Banerjee}, D. 2018, \apj, 853, 134,
	\dodoi{10.3847/1538-4357/aaa1a3}
	
	\bibitem[{{Mandal} {et~al.}(2016){Mandal}, {Magyar}, {Yuan}, {Van
			Doorsselaere}, \& {Banerjee}}]{2016ApJ...820...13M}
	{Mandal}, S., {Magyar}, N., {Yuan}, D., {Van Doorsselaere}, T., \& {Banerjee},
	D. 2016, \apj, 820, 13, \dodoi{10.3847/0004-637X/820/1/13}
	
	\bibitem[{{Mariska}(2006)}]{2006ApJ...639..484M}
	{Mariska}, J.~T. 2006, \apj, 639, 484, \dodoi{10.1086/499296}
	
	\bibitem[{{Mendoza-Brice{\~n}o} {et~al.}(2004){Mendoza-Brice{\~n}o},
		{Erd{\'e}lyi}, \& {Sigalotti}}]{2004ApJ...605..493M}
	{Mendoza-Brice{\~n}o}, C.~A., {Erd{\'e}lyi}, R., \& {Sigalotti}, L. D.~G. 2004,
	\apj, 605, 493, \dodoi{10.1086/382182}
	
	\bibitem[{{M{\'e}sz{\'a}rosov{\'a}} {et~al.}(2014){M{\'e}sz{\'a}rosov{\'a}},
		{Karlick{\'y}}, {Jel{\'\i}nek}, \& {Ryb{\'a}k}}]{2014ApJ...788...44M}
	{M{\'e}sz{\'a}rosov{\'a}}, H., {Karlick{\'y}}, M., {Jel{\'\i}nek}, P., \&
	{Ryb{\'a}k}, J. 2014, \apj, 788, 44, \dodoi{10.1088/0004-637X/788/1/44}
	
	\bibitem[{{M{\'e}sz{\'a}rosov{\'a}} {et~al.}(2011){M{\'e}sz{\'a}rosov{\'a}},
		{Karlick{\'y}}, \& {Ryb{\'a}k}}]{2011SoPh..273..393M}
	{M{\'e}sz{\'a}rosov{\'a}}, H., {Karlick{\'y}}, M., \& {Ryb{\'a}k}, J. 2011,
	\solphys, 273, 393, \dodoi{10.1007/s11207-011-9794-6}
	
	\bibitem[{{Murawski} {et~al.}(2011){Murawski}, {Zaqarashvili}, \&
		{Nakariakov}}]{2011A&A...533A..18M}
	{Murawski}, K., {Zaqarashvili}, T.~V., \& {Nakariakov}, V.~M. 2011, \aap, 533,
	A18, \dodoi{10.1051/0004-6361/201116942}
	
	\bibitem[{{Nakariakov} {et~al.}(2017){Nakariakov}, {Afanasyev}, {Kumar}, \&
		{Moon}}]{2017ApJ...849...62N}
	{Nakariakov}, V.~M., {Afanasyev}, A.~N., {Kumar}, S., \& {Moon}, Y.-J. 2017,
	\apj, 849, 62, \dodoi{10.3847/1538-4357/aa8ea3}
	
	\bibitem[{{Nakariakov} {et~al.}(2004){Nakariakov}, {Arber}, {Ault},
		{Katsiyannis}, {Williams}, \& {Keenan}}]{2004MNRAS.349..705N}
	{Nakariakov}, V.~M., {Arber}, T.~D., {Ault}, C.~E., {et~al.} 2004, \mnras, 349,
	705, \dodoi{10.1111/j.1365-2966.2004.07537.x}
	
	\bibitem[{{Nakariakov} \& {Kolotkov}(2020)}]{2020ARA&A..58..441N}
	{Nakariakov}, V.~M., \& {Kolotkov}, D.~Y. 2020, \araa, 58, 441,
	\dodoi{10.1146/annurev-astro-032320-042940}
	
	\bibitem[{{Nakariakov} {et~al.}(2019){Nakariakov}, {Kosak}, {Kolotkov},
		{Anfinogentov}, {Kumar}, \& {Moon}}]{2019ApJ...874L...1N}
	{Nakariakov}, V.~M., {Kosak}, M.~K., {Kolotkov}, D.~Y., {et~al.} 2019, \apjl,
	874, L1, \dodoi{10.3847/2041-8213/ab0c9f}
	
	\bibitem[{{Nakariakov} {et~al.}(2024{\natexlab{a}}){Nakariakov}, {Zhong},
		{Kolotkov}, {Meadowcroft}, {Zhong}, \& {Yuan}}]{2024RvMPP...8...19N}
	{Nakariakov}, V.~M., {Zhong}, S., {Kolotkov}, D.~Y., {et~al.}
	2024{\natexlab{a}}, Reviews of Modern Plasma Physics, 8, 19,
	\dodoi{10.1007/s41614-024-00160-9}
	
	\bibitem[{{Nakariakov} {et~al.}(2024{\natexlab{b}}){Nakariakov}, {Zhong}, \&
		{Kolotkov}}]{2024MNRAS.531.4611N}
	{Nakariakov}, V.~M., {Zhong}, Y., \& {Kolotkov}, D.~Y. 2024{\natexlab{b}},
	\mnras, 531, 4611, \dodoi{10.1093/mnras/stae1483}
	
	\bibitem[{{Nechaeva} {et~al.}(2019){Nechaeva}, {Zimovets}, {Nakariakov}, \&
		{Goddard}}]{Nechaeva2019Catalog}
	{Nechaeva}, A., {Zimovets}, I.~V., {Nakariakov}, V.~M., \& {Goddard}, C.~R.
	2019, \apjs, 241, 31, \dodoi{10.3847/1538-4365/ab0e86}
	
	\bibitem[{{Nistic{\`o}} {et~al.}(2013){Nistic{\`o}}, {Nakariakov}, \&
		{Verwichte}}]{2013A&A...552A..57N}
	{Nistic{\`o}}, G., {Nakariakov}, V.~M., \& {Verwichte}, E. 2013, \aap, 552,
	A57, \dodoi{10.1051/0004-6361/201220676}
	
	\bibitem[{{Ofman} \& {Wang}(2002)}]{2002ApJ...580L..85O}
	{Ofman}, L., \& {Wang}, T. 2002, \apjl, 580, L85, \dodoi{10.1086/345548}
	
	\bibitem[{{Ofman} {et~al.}(2012){Ofman}, {Wang}, \&
		{Davila}}]{2012ApJ...754..111O}
	{Ofman}, L., {Wang}, T.~J., \& {Davila}, J.~M. 2012, \apj, 754, 111,
	\dodoi{10.1088/0004-637X/754/2/111}
	
	\bibitem[{{Owen} {et~al.}(2009){Owen}, {De Moortel}, \&
		{Hood}}]{2009A&A...494..339O}
	{Owen}, N.~R., {De Moortel}, I., \& {Hood}, A.~W. 2009, \aap, 494, 339,
	\dodoi{10.1051/0004-6361:200810828}
	
	\bibitem[{{Pant} {et~al.}(2017){Pant}, {Tiwari}, {Yuan}, \&
		{Banerjee}}]{2017ApJ...847L...5P}
	{Pant}, V., {Tiwari}, A., {Yuan}, D., \& {Banerjee}, D. 2017, \apjl, 847, L5,
	\dodoi{10.3847/2041-8213/aa880f}
	
	\bibitem[{{Pascoe} {et~al.}(2017){Pascoe}, {Anfinogentov}, {Nistic{\`o}},
		{Goddard}, \& {Nakariakov}}]{2017A&A...600A..78P}
	{Pascoe}, D.~J., {Anfinogentov}, S., {Nistic{\`o}}, G., {Goddard}, C.~R., \&
	{Nakariakov}, V.~M. 2017, \aap, 600, A78, \dodoi{10.1051/0004-6361/201629702}
	
	\bibitem[{{Pascoe} {et~al.}(2016){Pascoe}, {Goddard}, {Nistic{\`o}},
		{Anfinogentov}, \& {Nakariakov}}]{2016A&A...589A.136P}
	{Pascoe}, D.~J., {Goddard}, C.~R., {Nistic{\`o}}, G., {Anfinogentov}, S., \&
	{Nakariakov}, V.~M. 2016, \aap, 589, A136,
	\dodoi{10.1051/0004-6361/201628255}
	
	\bibitem[{{Pascoe} {et~al.}(2012){Pascoe}, {Hood}, {de Moortel}, \&
		{Wright}}]{2012A&A...539A..37P}
	{Pascoe}, D.~J., {Hood}, A.~W., {de Moortel}, I., \& {Wright}, A.~N. 2012,
	\aap, 539, A37, \dodoi{10.1051/0004-6361/201117979}
	
	\bibitem[{{Prasad} {et~al.}(2021){Prasad}, {Srivastava}, \&
		{Wang}}]{2021SoPh..296..105P}
	{Prasad}, A., {Srivastava}, A.~K., \& {Wang}, T. 2021, \solphys, 296, 105,
	\dodoi{10.1007/s11207-021-01846-w}
	
	\bibitem[{{Prasad} {et~al.}(2022){Prasad}, {Srivastava}, {Wang}, \&
		{Sangal}}]{2022SoPh..297....5P}
	{Prasad}, A., {Srivastava}, A.~K., {Wang}, T., \& {Sangal}, K. 2022, \solphys,
	297, 5, \dodoi{10.1007/s11207-021-01940-z}
	
	\bibitem[{{Reale}(2014)}]{2014LRSP...11....4R}
	{Reale}, F. 2014, Living Reviews in Solar Physics, 11, 4,
	\dodoi{10.12942/lrsp-2014-4}
	
	\bibitem[{{Reale}(2016)}]{2016ApJ...826L..20R}
	---. 2016, \apjl, 826, L20, \dodoi{10.3847/2041-8205/826/2/L20}
	
	\bibitem[{{Ruderman}(2013)}]{2013A&A...553A..23R}
	{Ruderman}, M.~S. 2013, \aap, 553, A23, \dodoi{10.1051/0004-6361/201321175}
	
	\bibitem[{{Ruderman} {et~al.}(2025){Ruderman}, {Petrukhin}, \&
		{Kataeva}}]{2025MNRAS.542.1076R}
	{Ruderman}, M.~S., {Petrukhin}, N.~S., \& {Kataeva}, L.~Y. 2025, \mnras, 542,
	1076, \dodoi{10.1093/mnras/staf1310}
	
	\bibitem[{{Ruderman} \& {Roberts}(2002)}]{2002ApJ...577..475R}
	{Ruderman}, M.~S., \& {Roberts}, B. 2002, \apj, 577, 475,
	\dodoi{10.1086/342130}
	
	\bibitem[{{Sakurai} {et~al.}(2002){Sakurai}, {Ichimoto}, {Raju}, \&
		{Singh}}]{2002SoPh..209..265S}
	{Sakurai}, T., {Ichimoto}, K., {Raju}, K.~P., \& {Singh}, J. 2002, \solphys,
	209, 265, \dodoi{10.1023/A:1021297313448}
	
	\bibitem[{{Schrijver} {et~al.}(2015){Schrijver}, {Kauristie}, {Aylward},
		{Denardini}, {Gibson}, {Glover}, {Gopalswamy}, {Grande}, {Hapgood},
		{Heynderickx}, {Jakowski}, {Kalegaev}, {Lapenta}, {Linker}, {Liu},
		{Mandrini}, {Mann}, {Nagatsuma}, {Nandy}, {Obara}, {Paul O'Brien}, {Onsager},
		{Opgenoorth}, {Terkildsen}, {Valladares}, \& {Vilmer}}]{2015AdSpR..55.2745S}
	{Schrijver}, C.~J., {Kauristie}, K., {Aylward}, A.~D., {et~al.} 2015, Advances
	in Space Research, 55, 2745, \dodoi{10.1016/j.asr.2015.03.023}
	
	\bibitem[{{Selwa} {et~al.}(2005){Selwa}, {Murawski}, \&
		{Solanki}}]{2005A&A...436..701S}
	{Selwa}, M., {Murawski}, K., \& {Solanki}, S.~K. 2005, \aap, 436, 701,
	\dodoi{10.1051/0004-6361:20042319}
	
	\bibitem[{{Sigalotti} {et~al.}(2007){Sigalotti}, {Mendoza-Brice{\~n}o}, \&
		{Luna-Cardozo}}]{2007SoPh..246..187S}
	{Sigalotti}, L. D.~G., {Mendoza-Brice{\~n}o}, C.~A., \& {Luna-Cardozo}, M.
	2007, \solphys, 246, 187, \dodoi{10.1007/s11207-007-9077-4}
	
	\bibitem[{{Srivastava} \& {Goossens}(2013)}]{2013ApJ...777...17S}
	{Srivastava}, A.~K., \& {Goossens}, M. 2013, \apj, 777, 17,
	\dodoi{10.1088/0004-637X/777/1/17}
	
	\bibitem[{{Van Doorsselaere} {et~al.}(2021){Van Doorsselaere}, {Goossens},
		{Magyar}, {Ruderman}, \& {Ismayilli}}]{2021ApJ...910...58V}
	{Van Doorsselaere}, T., {Goossens}, M., {Magyar}, N., {Ruderman}, M.~S., \&
	{Ismayilli}, R. 2021, \apj, 910, 58, \dodoi{10.3847/1538-4357/abe630}
	
	\bibitem[{{Van Doorsselaere} {et~al.}(2024){Van Doorsselaere}, {Krishna
			Prasad}, {Pant}, {Banerjee}, \& {Hood}}]{2024A&A...683A.109V}
	{Van Doorsselaere}, T., {Krishna Prasad}, S., {Pant}, V., {Banerjee}, D., \&
	{Hood}, A. 2024, \aap, 683, A109, \dodoi{10.1051/0004-6361/202347579}
	
	\bibitem[{{Van Doorsselaere} {et~al.}(2011){Van Doorsselaere}, {Wardle}, {Del
			Zanna}, {Jansari}, {Verwichte}, \& {Nakariakov}}]{2011ApJ...727L..32V}
	{Van Doorsselaere}, T., {Wardle}, N., {Del Zanna}, G., {et~al.} 2011, \apjl,
	727, L32, \dodoi{10.1088/2041-8205/727/2/L32}
	
	\bibitem[{{Van Doorsselaere} {et~al.}(2020){Van Doorsselaere}, {Srivastava},
		{Antolin}, {Magyar}, {Vasheghani Farahani}, {Tian}, {Kolotkov}, {Ofman},
		{Guo}, {Arregui}, {De Moortel}, \& {Pascoe}}]{2020SSRv..216..140V}
	{Van Doorsselaere}, T., {Srivastava}, A.~K., {Antolin}, P., {et~al.} 2020,
	\ssr, 216, 140, \dodoi{10.1007/s11214-020-00770-y}
	
	\bibitem[{{Verwichte} {et~al.}(2008){Verwichte}, {Haynes}, {Arber}, \&
		{Brady}}]{2008ApJ...685.1286V}
	{Verwichte}, E., {Haynes}, M., {Arber}, T.~D., \& {Brady}, C.~S. 2008, \apj,
	685, 1286, \dodoi{10.1086/591077}
	
	\bibitem[{{Verwichte} {et~al.}(2013){Verwichte}, {Van Doorsselaere}, {White},
		\& {Antolin}}]{2013A&A...552A.138V}
	{Verwichte}, E., {Van Doorsselaere}, T., {White}, R.~S., \& {Antolin}, P. 2013,
	\aap, 552, A138, \dodoi{10.1051/0004-6361/201220456}
	
	\bibitem[{{Wang}(2011)}]{2011SSRv..158..397W}
	{Wang}, T. 2011, \ssr, 158, 397, \dodoi{10.1007/s11214-010-9716-1}
	
	\bibitem[{{Wang} \& {Ofman}(2019)}]{2019ApJ...886....2W}
	{Wang}, T., \& {Ofman}, L. 2019, \apj, 886, 2, \dodoi{10.3847/1538-4357/ab478f}
	
	\bibitem[{{Wang} {et~al.}(2015){Wang}, {Ofman}, {Sun}, {Provornikova}, \&
		{Davila}}]{2015ApJ...811L..13W}
	{Wang}, T., {Ofman}, L., {Sun}, X., {Provornikova}, E., \& {Davila}, J.~M.
	2015, \apjl, 811, L13, \dodoi{10.1088/2041-8205/811/1/L13}
	
	\bibitem[{{Wang} {et~al.}(2018){Wang}, {Ofman}, {Sun}, {Solanki}, \&
		{Davila}}]{2018ApJ...860..107W}
	{Wang}, T., {Ofman}, L., {Sun}, X., {Solanki}, S.~K., \& {Davila}, J.~M. 2018,
	\apj, 860, 107, \dodoi{10.3847/1538-4357/aac38a}
	
	\bibitem[{{Wang} {et~al.}(2021){Wang}, {Ofman}, {Yuan}, {Reale}, {Kolotkov}, \&
		{Srivastava}}]{2021SSRv..217...34W}
	{Wang}, T., {Ofman}, L., {Yuan}, D., {et~al.} 2021, \ssr, 217, 34,
	\dodoi{10.1007/s11214-021-00811-0}
	
	\bibitem[{{Wang} {et~al.}(2003{\natexlab{a}}){Wang}, {Solanki}, {Curdt},
		{Innes}, {Dammasch}, \& {Kliem}}]{2003A&A...406.1105W}
	{Wang}, T.~J., {Solanki}, S.~K., {Curdt}, W., {et~al.} 2003{\natexlab{a}},
	\aap, 406, 1105, \dodoi{10.1051/0004-6361:20030858}
	
	\bibitem[{{Wang} {et~al.}(2003{\natexlab{b}}){Wang}, {Solanki}, {Innes},
		{Curdt}, \& {Marsch}}]{2003A&A...402L..17W}
	{Wang}, T.~J., {Solanki}, S.~K., {Innes}, D.~E., {Curdt}, W., \& {Marsch}, E.
	2003{\natexlab{b}}, \aap, 402, L17, \dodoi{10.1051/0004-6361:20030448}
	
	\bibitem[{{Yuan} {et~al.}(2015){Yuan}, {Van Doorsselaere}, {Banerjee}, \&
		{Antolin}}]{2015ApJ...807...98Y}
	{Yuan}, D., {Van Doorsselaere}, T., {Banerjee}, D., \& {Antolin}, P. 2015,
	\apj, 807, 98, \dodoi{10.1088/0004-637X/807/1/98}
	
	\bibitem[{Zavershinskii \& Agapova(2026)}]{Zavershinskii2026}
	Zavershinskii, D., \& Agapova, D. 2026, Solar Physics, 301,
	\dodoi{10.1007/s11207-025-02602-0}
	
	\bibitem[{{Zavershinskii} {et~al.}(2021){Zavershinskii}, {Kolotkov},
		{Riashchikov}, \& {Molevich}}]{2021SoPh..296...96Z}
	{Zavershinskii}, D., {Kolotkov}, D., {Riashchikov}, D., \& {Molevich}, N. 2021,
	\solphys, 296, 96, \dodoi{10.1007/s11207-021-01841-1}
	
	\bibitem[{{Zavershinskii} {et~al.}(2023){Zavershinskii}, {Molevich},
		{Riashchikov}, \& {Belov}}]{2023FrASS..1067781Z}
	{Zavershinskii}, D.~I., {Molevich}, N.~E., {Riashchikov}, D.~S., \& {Belov},
	S.~A. 2023, Frontiers in Astronomy and Space Sciences, 10, 1167781,
	\dodoi{10.3389/fspas.2023.1167781}
	
	\bibitem[{{Zhao} {et~al.}(2025){Zhao}, {Wang}, \& {Chen}}]{2025MNRAS.538..797Z}
	{Zhao}, J., {Wang}, T., \& {Chen}, R. 2025, \mnras, 538, 797,
	\dodoi{10.1093/mnras/staf309}
	
	\bibitem[{{Zhong} {et~al.}(2023){Zhong}, {Kolotkov}, {Zhong}, \&
		{Nakariakov}}]{2023MNRAS.525.5033Z}
	{Zhong}, Y., {Kolotkov}, D.~Y., {Zhong}, S., \& {Nakariakov}, V.~M. 2023,
	\mnras, 525, 5033, \dodoi{10.1093/mnras/stad2598}
	
	\bibitem[{{Zimovets} {et~al.}(2021){Zimovets}, {McLaughlin}, {Srivastava},
		{Kolotkov}, {Kuznetsov}, {Kupriyanova}, {Cho}, {Inglis}, {Reale}, {Pascoe},
		{Tian}, {Yuan}, {Li}, \& {Zhang}}]{2021SSRv..217...66Z}
	{Zimovets}, I.~V., {McLaughlin}, J.~A., {Srivastava}, A.~K., {et~al.} 2021,
	\ssr, 217, 66, \dodoi{10.1007/s11214-021-00840-9}
	
\end{thebibliography}



\end{document}